\documentclass[a4paper,12pt]{article}
\usepackage{amsfonts,slashed}
\usepackage{latexsym}
\usepackage{amsfonts}
\usepackage{amsmath}
  \usepackage{amsmath,amssymb,amsthm}
\usepackage{slashed}
\usepackage{arydshln}

\makeatletter
\g@addto@macro\bfseries{\boldmath}
\makeatother

\usepackage{cite}

\DeclareFontFamily{OMS}{rsfs}{\skewchar\font'60}
\DeclareFontShape{OMS}{rsfs}{m}{n}{<-5>rsfs5 <5-7>rsfs7 <7->rsfs10 }{}
\DeclareSymbolFont{rsfs}{OMS}{rsfs}{m}{n}
\DeclareSymbolFontAlphabet{\Scr}{rsfs}

\newcommand\ocirc[1]{\ThisStyle{\ensurestackMath{%
  \stackon[1pt]{\SavedStyle#1}{\SavedStyle\kern.6\LMpt\circ}}}}
\numberwithin{equation}{section}
\def\be{\begin{equation}}
\def\ee{\end{equation}}
\def\ba{\begin{array}}
\def\ea{\end{array}}

\newcommand{\bea}{\begin{eqnarray}}
\newcommand{\eea}{\end{eqnarray}}

\def\Re{{\rm Re}}
\def\Im{{\rm Im}}

\def\={~=~}
\def\*{{}^*}

\def\SO{\mathrm{SO}}

\def\={~=~}
\def\*{{}^*}

\def\SO{\mathrm{SO}}

\begin{document}

\begin{titlepage}

\rightline{December 2016}
\begin{center}
\vskip 1.0cm
{\LARGE \bf Type II origin of dyonic gaugings}\\
\vskip 1.0cm

{\large\bf {Gianluca Inverso${\,}^1$, Henning Samtleben${\,}^2$, Mario Trigiante${\,}^3$}}
\vskip .6cm
{\it ${}^1$
Center for Mathematical Analysis, Geometry and Dynamical Systems, \\
Department of Mathematics, Instituto Superior T\'ecnico,\\
Universidade de Lisboa, Av. Rovisco Pais, 1049-001 Lisboa, Portugal}\\
ginverso@math.tecnico.ulisboa.pt
\vskip .2cm

{\it ${}^2$
Univ Lyon, Ens de Lyon, Univ Claude Bernard, CNRS,\\
Laboratoire de Physique, F-69342 Lyon, France} \\
{henning.samtleben@ens-lyon.fr}
\vskip .2cm

{\it ${}^3$ Dipartimento di Fisica, Politecnico di Torino\\
C.~so Duca degli Abruzzi, 24, I-10129 Torino, Italy\\
and Istituto Nazionale di Fisica Nucleare (INFN) 
Sezione di Torino,\\ Via Pietro Giuria 1, 10125 Torino, Italy} \\
mario.trigiante@polito.it

\vskip 1cm

\vskip 1cm
{\bf Abstract}
\end{center}


\begin{narrower}

\noindent
Dyonic gaugings of four-dimensional supergravity typically exhibit a richer vacuum structure
compared to their purely electric counterparts,
but their higher-dimensional origin often remains more mysterious.
We consider a class of dyonic gaugings with gauge groups of the
type $({\rm SO}(p,q)\times {\rm SO}(p',q')) \ltimes N$ with $N$ nilpotent.
Using generalized Scherk-Schwarz reductions of exceptional field theory,
we show how these four-dimensional gaugings may be consistently
embedded in Type II supergravity upon compactification around products of spheres and hyperboloids.
As an application, we give the explicit uplift of the ${\cal N}=4$ AdS$_4$ vacuum of the theory
with gauge group $({\rm SO}(6)\times {\rm SO}(1,1))\ltimes T^{12}$ into a
supersymmetric AdS$_4\times M_5\times S^1$ S-fold solution of IIB supergravity. The internal space
$M_5$ is a squashed $S^5$ preserving an ${\rm SO}(4)\subset {\rm SO}(6)$ subset of its isometries.

\end{narrower}

\end{titlepage}
\newpage

\tableofcontents

\section{Introduction}

Ungauged supergravities in four space-time dimensions are defined up to a choice of the electric-magnetic symplectic frame. Different frames yield physically equivalent ungauged models, though described by inequivalent Lagrangians.
Things change when the theory is gauged, namely when a suitable global symmetry group of the ungauged Lagrangian is promoted to local symmetry through the so-called gauging procedure.
In the presence of extended supersymmetry, only gauged supergravities, as opposed to their ungauged counterparts, can have a scalar potential, and thus a non-trivial vacuum structure.  When constructing these models through the gauging procedure, the initial choice of the symplectic frame becomes physically relevant. This freedom can be taken into account by allowing, in a fixed symplectic frame, for magnetic components of the embedding tensor defining the gauge algebra, namely by considering \emph{dyonic gaugings}.\footnote{All the gaugings we consider, including `dyonic' ones, satisfy locality constraints. Namely, there always exists a choice of symplectic frame in which the gauging is entirely electric.}
Different initial frames will in general yield different choices of the gauge group and even gauging a same group in different frames may yield physically distinct theories.

This feature was first exploited in ${\cal N}=4$ supergravity \cite{deRoo:1985jh}.
In the maximal theory the freedom in the initial choice of symplectic frame led to the discovery of new gaugings in \cite{Andrianopoli:2002mf,deWit:2002vt,Hull:2002cv,deWit:2007mt} and, more recently, in ~\cite{DallAgata:2011aa,DallAgata:2012bb,DallAgata:2012sx,DallAgata:2014ita}.
In \cite{DallAgata:2012bb,DallAgata:2012sx}, in particular,
one-parameter families of dyonic ${\rm SO}(p,q)$-gaugings were found in $\mathcal{N}=8,\,D=4$ supergravity, generalizing their well known electric counterparts \cite{deWit:1982bul,Hull:1984vg}. These new models were constructed by gauging the same ${\rm SO}(p,q)$ group in different frames, the choice of which is parametrized by a continuous angular parameter~$\omega$. They are known as $\omega$-deformed models, where the value $\omega=0$ corresponds to the original electric gaugings of \cite{deWit:1982bul,Hull:1984vg}. The parameter $\omega$ is physical in that its value can not be offset by field redefinitions or the action of the global symmetry group $G$ of the ungauged theory, and does affect the physics of the model.\par A different class of dyonic models, originally devised in \cite{DallAgata:2011aa}, are based on non-semisimple groups of the form
\begin{equation}
\left({\rm SO}(p,q) \times {\rm SO}(p',q')\right) \ltimes N
 \;,
\end{equation}
with $p+q+p'+q'\le8$ and $N$ is a subgroup generated by a nilpotent algebra whose properties are described later. These gauge groups can be defined as different contractions of the semisimple group ${\rm SO}(p+p',8-p-p')$, generalizing the ${\rm CSO}(p,q,r)$-gaugings of \cite{Hull:1984vg}.
They are characterized by a ${\rm CSO}(p,q,8-p-q)$ subgroup gauged by the electric vector fields and a ${\rm CSO}(p',q',8-p'-q')$ gauged by the magnetic ones, with a subset of the nilpotent generators gauged by a combination of the two fields.
As opposed to the $\omega$-deformed ${\rm SO}(p,q)$-models, the corresponding gauged theories, also known also as \emph{dyonic} ${\rm CSO}(p,q,r)$ models, do not depend on a continuous parameter aside from an overall coupling constant. The only exceptions are the ${\rm SO}(4)^2\ltimes\mathbb R^{16}$ gaugings and their non-compact forms, which have a one-parameter family of deformations corresponding to the ratio of gauge couplings for the two semisimple factors \cite{DallAgata:2014ita}.
\par
Dyonic gaugings feature a richer vacuum structure than their original electric counterparts. Of particular interest, for their application to the AdS/CFT correspondence, are the anti-de Sitter vacua. In order to understand the general features of the dual three-dimensional CFT, however, a UV completion of the model within superstring or M-theory is called for. The original ${\rm SO}(8)$-gauged maximal supergravity of \cite{deWit:1982bul} features a maximally supersymmetric AdS vacuum and describes a consistent truncation of eleven-dimensional supergravity compactified on a seven-sphere \cite{deWit:1986oxb}. The CFT dual to the maximally supersymmetric vacuum of the theory is the ABJM model \cite{Aharony:2008ug}.  The electric ${\rm CSO}(p,q,r)$-gaugings, on the other hand, describe consistent truncations of eleven-dimensional theories on backgrounds in which the internal manifold has the form $H^{p,q}\times \mathbb{R}^r$, $H^{p,q}$ being a hyperboloid \cite{Hull:1988jw,Hohm:2014qga}.
As for the dyonic gaugings, while the ten or eleven-dimensional origin of the $\omega$-deformed ${\rm SO}(p,q)$-models is as yet elusive \cite{deWit:2013ija,Lee:2015xga}, some progress has been made for the dyonic ${\rm CSO}(p,q,r)$-supergravities: Recently
the dyonic ${\rm ISO}(7)$-model was interpreted as a consistent truncation of massive Type IIA string theory \cite{Romans:1985tz} on a background with topology of the form $AdS_4\times S^6$ \cite{Guarino:2015jca,Guarino:2015qaa,Guarino:2015vca}. In the present work we make further progress in this direction by defining a ten-dimensional origin for all the remaining dyonic ${\rm CSO}(p,q,r)$-models. Of special interest is the dyonic-model with gauge group $\left({\rm SO}(6)\times {\rm SO}(1,1)\right)\ltimes T^{12}$, which features a characteristic $\mathcal{N}=4$ AdS vacuum \cite{Gallerati:2014xra} of which we give a ten-dimensional description in the Type IIB theory. \par
Exceptional field theory (ExFT) \cite{Hohm:2013pua,Hohm:2013uia,Hohm:2014qga} has proven to be a valuable framework to study the higher-dimensional origin of $D$-dimensional maximal gauged theories. It provides a formulation of maximal supergravities, including the eleven and the ten-dimensional ones, which is manifestly covariant with respect to the on-shell global symmetry group of the $D$-dimensional model. In our analysis we are interested in uplifting four-dimensional maximal gauged supergravities \cite{deWit:2007mt} so we choose to work in the $D=4$ formulation of ExFT in which the manifest duality symmetry is the ${\rm E}_{7(7)}$ on-shell invariance of the Cremmer-Julia ungauged four-dimensional $\mathcal{N}=8$ theory \cite{Cremmer:1979up}. In this framework the fields of the
 $D=4$, ${\cal N}=8$ supergravity are described as formally depending, in addition to the four space-time coordinates $x^\mu$, on $56$ coordinates $Y^M$ in the fundamental representation of ${\rm E}_{7(7)}$. This dependence is strongly restricted by the so-called \emph{section-constraints} \cite{Coimbra:2011ky,Berman:2012vc}. Solutions to these constraints describe the eleven and ten-dimensional massless maximal supergravities written in terms of $D=4$ fields, which only depend on specific sets of seven and six internal coordinates, respectively.
In \cite{Ciceri:2016dmd} a deformed version of ExFT was defined in order to describe the massive Type IIA theory and its consistent truncations to $D=4$.\footnote{See \cite{Hohm:2011cp} and \cite{Cassani:2016ncu} for corresponding results in the contexts of Double Field Theory and Exceptional Generalized Geometry, respectively.}\par
The embedding of a gauged four-dimensional model in the eleven or ten-dimensional theories is effected through a suitable Scherk-Schwarz ansatz \cite{Hohm:2014qga} in which the ExFT fields depend on the internal coordinates through an ${\rm E}_{7(7)}$-valued \emph{twist matrix} $U_{M}{}^{{N}}(Y)$. This matrix encodes the higher dimensional fields as well as the fluxes on a certain background around which the four-dimensional fields ought to describe fluctuations.
 For instance the  Scherk-Schwarz ansatz for the scalar fields of the ExFT is written in terms of the characteristic symmetric symplectic ${\rm E}_{7(7)}$-matrix ${\cal M}_{MN}(x,Y)$ as follows:
 \bea
{\cal M}_{MN}(x,Y) &=& U_{M}{}^{{K}}(Y)\,U_{N}{}^{{L}}(Y)\,M_{{KL}}(x)\;,
\label{genSS_intro}
\eea
where $M_{{KL}}(x)$ describes $D=4$ scalar fluctuations about the higher-dimensional background whose fields (metric, form-fields and fluxes) are encoded in the matrix $U_{M}{}^{{N}}(Y)$.
   If certain conditions on the twist matrix are satisfied, the dependence of the fields on the internal coordinates through $U(Y)$ factors out in the ExFT field equations, yielding the field equations of gauged four-dimensional model in the $x^\mu$-dependent fields.
 The corresponding embedding tensor is encoded in $U_{M}{}^{{N}}(Y)$. The section constraints restrict the $Y$-dependence of this matrix and thus the possible gauged models which can be described as consistent truncations of the ten or eleven dimensional theories.\par
 In the present paper the embedding of the dyonic ${\rm CSO}(p,q,r)$-gaugings, with $p+q\ge 2, r\ge 2$, in the Type II theories is effected by writing the twist matrix $U(Y)$ as the product of two commuting matrices $\hat{U}(y^i)$ and $ \mathring{U}(\tilde{y}_a)$;
 \begin{equation}
 U(y^i,\,\tilde{y}_a)=\hat{U}(y^i)\mathring{U}(\tilde{y}_a)\,\,,\,\,\,\,i=1,\dots,\,p+q-1\,\,,\,\,a=p+q,\dots,\,6\,,
 \end{equation}
These two matrices separately define the electric $\mathfrak{cso}(p,q,r)$ and the magnetic $\mathfrak{cso}(p',q',r')$ subalgebras and the corresponding sets of coordinates $\{y^i\}$ and $\{\tilde{y}_a\}$ are chosen within distinct ${\rm SL}(8)$ representations satisfying a suitable condition of mutual compatibility. The total twist matrix satisfies the section constraints so that the corresponding dyonic models can be embedded either in Type IIA ($p+q$ odd) or in Type IIB ($p+q$ even) theories. \par
The dyonic model with $p=6,\,q=0,\,p'=q'=1$ mentioned earlier corresponds to a gauge group of the form $\left({\rm SO}(6)\times {\rm SO}(1,1)\right)\ltimes T^{12}$. It can be obtained from a step-wise compactification of the Type IIB theory as follows. A first compactification of Type IIB on $AdS_5\times S^5$ yields five-dimensional supergravity with gauge group ${\rm SO}(6)$ \cite{Gunaydin:1985cu,Pernici:1985ju,Baguet:2015sma}. This model still features the ${\rm SL}(2,\mathbb{R})$ duality symmetry of the Type IIB theory,  commuting with ${\rm SO}(6)$. As a last step one can perform a Scherk-Schwarz reduction down to $D=4$, choosing a twist matrix valued in an ${\rm SO}(1,1)$ subgroup of ${\rm SL}(2,\mathbb{R})$. The resulting model supports the above mentioned AdS$_4$ vacuum (however not at the scalar origin) preserving $\mathcal{N}=4$ supersymmetries of which we give a Type IIB description. Its geometry is an AdS$_4 \times M_5 \times S^1$ S-fold with the internal space
$M_5$ given by a deformation of the round sphere $S^5$ preserving an ${\rm SO}(3)\times {\rm SO}(3)\subset {\rm SO}(6)$ subset of its isometries.

\par
The paper is organized as follows: In Section~\ref{s:2} we recall the main facts about the dyonic ${\rm CSO}(p,q,r)$ gaugings. Section~\ref{s:3} gives a brief review of the relevant ExFT.
In Sections \ref{s:4} and \ref{s:5} the Scherk-Schwarz ansaetze defining the Type II embedding of the dyonic ${\rm CSO}(p,q,r)$ gaugings are discussed in detail. Finally,
in Section \ref{s:6} we focus on the  $(\rm SO(6)\times SO(1,1))\ltimes T^{12}$ gauged maximal supergravity and work out, using the general ExFT description of  Type IIB theory and the corresponding Scherk-Schwarz ansatz, its uplift into the IIB theory.
In particular, we give the uplift of the four-dimensional $\mathcal{N}=4$ AdS vacuum into a IIB S-fold solution. In Appendix~\ref{janusmatch} we also prove that the non-compact version of this ten-dimensional geometry (i.e. before S-folding) falls in the class of Janus solutions found in \cite{DHoker:2007zhm,DHoker:2007hhe}. We end with some concluding remarks.

\section{Type II origin of dyonic gaugings}

\subsection{Dyonic gaugings}\label{s:2}

Gaugings of maximal $D=4$ supergravity are conveniently described by the embedding tensor formalism \cite{Cordaro:1998tx,Nicolai:2000sc,deWit:2002vt,deWit:2005ub,deWit:2007mt} (for reviews see \cite{Samtleben:2008pe,Trigiante:2016mnt}).

All the information about the gauge couplings of the theory is encoded into a tensor $X_{MN}{}^P$ transforming in the $\bf 912$ representation of E$_{7(7)}$, where indices $M,N,\ldots$ correspond to the $\bf56$ representation.
In an appropriate symplectic frame an SL(8) subgroup of E$_{7(7)}$ acts separately on electric and magnetic vectors.
We are interested in non-semisimple gauge groups
contained in SL$(8,\mathbb R)$ of the form
\begin{equation}
(\mathrm{SO}(p,q)\times\mathrm{SO}(p',q'))\ltimes N\,,
\label{dyonic gaugings}
\end{equation}
with $N$ a nilpotent factor which becomes abelian when $p+q+p'+q'=8$ \cite{DallAgata:2011aa} (see also \cite{Trigiante:2016mnt} for a review).
Its generators in the fundamental of SL(8) are triangular matrices with non-vanishing entries either in the first $p+q$ rows and last $8-p-q$ columns, or in the first $8-p'-q'$ columns and last $p'+q'$ rows. These two sets of nilpotent generators overlap on a common $(p+q)(p'+q')$-dimensional abelian subalgebra.
This class of gaugings is described by two symmetric matrices $\eta_{AB}$, $\tilde\eta^{AB}$ corresponding to the $\bf36'$ and $\bf36$ irreps in the decomposition of the $\bf912$ under SL(8), with $A,B,\ldots$ fundamental SL(8) indices.
Up to SL$(8)$ transformations we can write%
\footnote{If we take instead $\eta_{AB}$ invertible and $\tilde\eta^{AB}\propto(\eta_{AB})^{-1}$, the resulting gaugings are the families of $\omega$-deformed SO($p,q$) gauged maximal supergravities \cite{DallAgata:2012bb}.}
\begin{equation}
\begin{split}
\eta_{AB} &= \mathrm{diag}(\overbrace{1,\ldots,1}^{p},\overbrace{-1,\ldots,-1}^{q},0,\ldots\ldots,0)\,,\\
\tilde\eta^{AB} &\propto \mathrm{diag}(0,\ldots\ldots,0,\underbrace{1,\ldots,1}_{p'},\underbrace{-1,\ldots,-1}_{q'})\,,
\end{split}
\end{equation}
such that $\eta_{AC}\tilde\eta^{CB}=0$ in order to solve the embedding tensor quadratic constraints \cite{DallAgata:2011aa}.
The embedding tensor then takes the form
\bea
X_{AB,CD}{}^{EF} &=& \eta_{A[C} \delta_{D]B}{}^{EF} - \eta_{B[C} \delta_{D]A}{}^{EF} \;,
\nonumber\\
X^{AB}{}_{CD}{}^{EF} &=& -\tilde\eta^{A[E} \delta_{CD}{}^{F]B} + \tilde\eta^{B[E} \delta_{CD}{}^{F]A} \;,
\label{embeddingX}
\eea
where $\bf56$ E$_{7(7)}$ indices $M,N,\ldots$ are decomposed into the ${\bf28}'+{\bf28}$ of SL$(8,\mathbb R)$, described by upper and lower antisymmetrized pairs of $\bf8$ indices.

Most of these gauged models are entirely specified by their gauge group embedded in SL(8), with a few notable exceptions \cite{DallAgata:2014ita}.
When $p+q=7$ we find ISO$(p,q)$ gaugings.
In this case the gauge group is entirely specified by $\eta_{AB}$ and  $\tilde\eta^{AB}$ only affects the gauge connection of the $\mathbb R^7$ subgroup.
A non-vanishing $\tilde\eta^{AB}$ is identified with the Romans mass in a IIA uplift of the gauging \cite{Guarino:2015jca,Guarino:2015qaa,Guarino:2015vca}.
Moreover, when $p+q=p'+q'=4$ the relative overall normalization of $\tilde\eta^{AB}$ with respect to $\eta_{AB}$ cannot be reabsorbed in any E$_{7(7)}$ transformation and thus determines a one-parameter family of inequivalent gaugings sharing the same gauge group.

Several of the dyonic CSO$(p,q,r)$ models exhibit interesting vacua.
Maximally symmetric vacuum solutions of the resulting gauged maximal supergravities are determined by extrema of the scalar potential \cite{deWit:2007mt}
\bea
V(\phi) &=&
  \frac1{672}\,  {M}(\phi)^{MP} \left(X_{MN}{}^{R}X_{PQ}{}^{S}{M}(\phi)^{NQ}{ M}(\phi)_{RS} + 7\,X_{MN}{}^{Q}X_{PQ}{}^{N} \right)
  \;,
\label{sugra scal pot}
\eea
where $ M(\phi)_{MN}$ is a symmetric matrix parameterizing the E$_{7(7)}/\mathrm{SU(8)}$ non-linear sigma model of the scalar fields, and $M(\phi)^{MN}$ is its inverse.
The deformed ISO(7) gauging (i.e. with $\tilde\eta\neq0$) has several supersymmetric and non-supersymmetric AdS$_4$ solutions \cite{DallAgata:2011aa,Borghese:2012qm,Gallerati:2014xra}.
The $(\mathrm{SO}(4)\times \mathrm{SO}(2,2))\ltimes T^{16}$ gauging (with equal normalizations for $\eta_{AB}$ and $\tilde\eta^{AB}$) and the $(\mathrm{SO(2)\times SO}(2))\ltimes N_{20}$ model are part of a large class of theories exhibiting Minkowski vacua connected through singular limits in their moduli spaces \cite{Catino:2013ppa}.
In this paper we will focus in particular on the $(\mathrm{SO(6)\times SO(1,1)})\ltimes T^{12}$ gauging which is known to have an $\mathcal N=4$ AdS$_4$ vacuum \cite{Gallerati:2014xra}, in addition to other unstable AdS$_4$ solutions~\cite{DallAgata:2011aa}.

\subsection{Scherk-Schwarz reduction in exceptional field theory}\label{s:3}

Exceptional field theories are the manifestly duality covariant reformulations
of maximal supergravities. Since our goal is a higher-dimensional embedding of
four-dimensional maximal supergravities, which are obtained as gaugings~\cite{deWit:2007mt}
of the ${\rm E}_{7(7)}$-invariant Cremmer-Julia theory~\cite{Cremmer:1979up},
the proper framework for their higher-dimensional embedding is the
${\rm E}_{7(7)}$-covariant exceptional field theory constructed in~\cite{Hohm:2013uia}.
This exceptional field theory is formulated in terms of the fields of $D=4$, ${\cal N}=8$ supergravity
which in addition to the $4$ external coordinates $x^\mu$ formally depend on
$56$ internal coordinates $Y^M$ forming the fundamental representation of ${\rm E}_{7(7)}$.
The latter dependence is however severely restricted by the section
constraints~\cite{Coimbra:2011ky,Berman:2012vc}
 \bea
\Omega^{MK}
 (t_\alpha)_K{}^{N}\,\partial_M \partial_N A =  0\;, \quad
 \Omega^{MK}(t_\alpha)_K{}^{N}\,\partial_MA\, \partial_N B =  0
  \;, \quad
  \Omega^{MN}\,\partial_MA\, \partial_N B = 0 \,.
 \label{sectioncondition}
 \eea
Here $(t_\alpha)^{MN}$ and $\Omega^{MN}$ denote the ${\rm E}_{7(7)}$ generators and the
symplectic invariant antisymmetric matrix, respectively. The section constraints
(\ref{sectioncondition}) admit two inequivalent solutions restricting the internal coordinate
dependence to a subset of coordinates, identified upon breaking ${\rm E}_{7(7)}$
down to ${\rm G}_{\rm IIA}={\rm GL}(6)$ and ${\rm G}_{\rm IIB}={\rm GL}(6)\times{\rm SL}(2)$, respectively
\bea
11D/{\rm IIA} ~:~
 {\bf 56} &\longrightarrow&
\boxed{{\mathbf6'}_{-4}+ {\mathbf1}_{-3}}+{\mathbf6}_{-2}+{\mathbf{15}}_{-1}+ \mathbf{15}'_{+1}+\mathbf6'_{+2}+1_{+3}+\mathbf6_{+4}\;,
\nonumber\\
{\rm IIB} ~:~{\bf 56} &\longrightarrow&
\boxed{({\mathbf6'},\mathbf1)_{-4}}+(\mathbf6,\mathbf2)_{-2}+(\mathbf{20},\mathbf1)_0 +(\mathbf6',\mathbf2)_{+2}+ (\mathbf6,\mathbf1)_{+4}
\;.
\label{coordinates}
\eea
The former solution allows for the dependence of all fields on $6+1$ coordinates, upon which
the field equations of exceptional field theory reduce to those of $D=11$ supergravity.
In the latter solution, fields depend on a maximal set of 6 coordinates which are singlet
under the ${\rm SL}(2)\subset {\rm G}_{\rm IIB}$. The resulting field equations thus exhibit a
global ${\rm SL}(2)$ symmetry and coincide with the equations of IIB supergravity. The decomposition
(\ref{coordinates}) shows that ${\rm G}_{\rm IIA}$ and ${\rm G}_{\rm IIB}$ intersect on a common ${\rm GL}(5)$\,.

For the details of the ${\rm E}_{7(7)}$ exceptional field theory, in particular
its Lagrangian and field equations, we refer to~\cite{Hohm:2013uia,Godazgar:2014nqa}.
Here we just review its bosonic field content
\bea
\{g_{\mu\nu}, {\cal M}_{MN}, {\cal A}_{\mu}{}^{M}, {\cal B}_{\mu\nu\,\alpha}, {\cal B}_{\mu\nu\,M}
\}
\;,
\label{fcontent}
\eea
drawing on the field content of $D=4$ maximal supergravity. The matrices $g_{\mu\nu}$
and ${\cal M}_{MN}$ represent the external and internal metric, respectively, with the latter
parametrizing the ${\rm E}_{7(7)}/{\rm SU}(8)$ coset space. The vectors ${\cal A}_{\mu}{}^{M}$
and two-forms $\{{\cal B}_{\mu\nu\,\alpha}, {\cal B}_{\mu\nu\,M}\}$ transform in the
${\bf 56}$ and ${\bf 133}\oplus{\bf 56}$ of ${\rm E}_{7(7)}$, respectively.
In order to establish the equivalence with IIA/IIB supergravity after solving
the section constraint, the fields (\ref{fcontent})
are decomposed w.r.t.\ the relevant ${\rm G}_{\rm IIA, IIB}$ defining
(\ref{coordinates}).
E.g.\ the scalar matrix ${\cal M}_{MN}$ is parametrized as ${\cal M}={\cal V}{\cal V}^{\rm T}$
in terms of the group-valued vielbein ${\cal V}$,
parametrized in the triangular gauge associated with the ${\rm GL}(1)\subset {\rm G}_{\rm IIA, IIB}$
grading according to \cite{Cremmer:1997ct}. In IIB parametrization this takes the form of an expansion
\bea
{\cal V}_{\rm IIB} &\equiv&
{\rm exp}\left[b_{\alpha}\,t_{(+3)}^{\alpha}\right]
{\rm exp}\left[\epsilon^{klmnpq}\,c_{klmn}\,t_{(+2)}{\,}_{pq}\right]
{\rm exp}\left[b_{mn}{}^{\alpha}\,t_{(+1)}^{mn}{}_{\alpha}\right]\,
{\cal V}_6\,{\cal V}_2\;
{\rm exp} \left[\phi\, t^{\rm IIB}_{(0)}\right]
\;,\quad
\label{V56}
\eea
in which one recognizes the various IIB fields.%
\footnote{Depending on the context, indices $\alpha,\beta,\ldots$ represent either the E$_{7(7)}$ adjoint representation or the SL(2) fundamental. This should cause no confusion.}
A similar expansion holds for the IIA parametrization.
The precise dictionary between the ExFT formulation and IIA/IIB
supergravity further requires redefinitions of all the form fields originating from the higher-dimensional $p$-forms
in the usual Kaluza-Klein manner, as well as a series of dualization and non-linear field redefinitions,
c.f.~\cite{Hohm:2013vpa,Hohm:2013uia}.

Consistent truncations in exceptional field theory are conveniently constructed via a
generalized Scherk-Schwarz reduction by the ansatz~\cite{Hohm:2014qga}
\bea\label{SSansatz}
 g_{\mu\nu}(x,Y) &=& \rho^{-2}(Y)\,g_{\mu\nu}(x)\;,\nonumber\\
 {\cal M}_{MN}(x,Y) &=& U_{M}{}^{{K}}(Y)\,U_{N}{}^{{L}}(Y)\,M_{{KL}}(x)\;,
\nonumber\\
  {\cal A}_{\mu}{}^{M}(x,Y) &=& \rho^{-1}(Y) A_{\mu}{}^{{N}}(x)(U^{-1})_{{N}}{}^{M}(Y) \;,
  \nonumber\\
  {\cal B}_{\mu\nu\,{\alpha}}(x,Y) &=& \,\rho^{-2}(Y) U_{{\alpha}}{}^{{\beta}}(Y)\,B_{\mu\nu\,{\beta}}(x)
  \;,
\nonumber\\
{\cal B}_{\mu\nu\,M}(x,Y) &=&
-2\, \rho^{-2}(Y)\,(U^{-1})_{{S}}{}^P(Y) \,\partial_M U_P{}^{{R}}(Y) (t^{{\alpha}}){}_{{R}}{}^{{S}}\, B_{\mu\nu\,{\alpha}}(x)
\;,
\eea
for the bosonic fields (\ref{fcontent}). The dependence on the internal coordinates is carried by
an ${\rm E}_{7(7)}$-valued twist matrix $U_M{}^{{N}}$ and a scale factor $\rho(Y)$,
satisfying the first order differential equations~\cite{Aldazabal:2013mya}
\bea
\left[(U^{-1})_{{M}}{}^{P} (U^{-1})_{{N}}{}^{Q} \, \partial_P U_Q{}^{{K}}\right]_{{\bf 912}}
&\stackrel{!}{=}&
\frac1{7}\,\rho \,\Theta_{{M}}{}^\alpha\, (t_\alpha)_{{N}}{}^{{K}}
\;,
\nonumber\\
  \partial_N (U^{-1})_{{M}}{}^N
- 3\,\rho^{-1}\partial_N \rho  \, (U^{-1})_{{M}}{}^N  &\stackrel{!}{=}&
2\,\rho\,\vartheta_{{M}}
\;,
\label{consistent_twist}
\eea
with constant tensors $\Theta_{{M}}{}^\alpha$ and $\vartheta_{{M}}$.
The latter can be identified with the irreducible components of the
embedding tensor of the four-dimensional gauged supergravity \cite{deWit:2007mt} to which the theory
reduces after the generalized Scherk-Schwarz ansatz. In particular, the notation $[\cdot]_{{\bf 912}}$ refers to projection
onto the irreducible ${\bf 912}$ representation of ${\rm E}_{7(7)}$.

 Every solution to the system (\ref{consistent_twist}) defines a consistent truncation of
 exceptional field theory down to a four-dimensional gauged supergravity with all $Y$-dependence
 consistently factoring out from the field equations. If the matrix $U_M{}^{{N}}$ and the scale factor $\rho(Y)$
 satisfy the section constraint (\ref{sectioncondition}), the dictionary with IIA/IIB supergravity provides
 the explicit formulas for a geometrical uplift of the resulting four-dimensional gauging into Type II
 supergravity.
 In this paper, we will construct the twist matrices $U_M{}^{{N}}$ that define the geometrical
 uplift of the dyonic gaugings defined above.

\subsection{Scherk-Schwarz twist matrices for dyonic gaugings}\label{s:4}

The solutions to the consistency equations  (\ref{consistent_twist}) constructed in \cite{Hohm:2014qga}
give rise to the embedding tensors associated with the gaugings of ${\rm SO}(p,q)$ and ${\rm CSO}(p,q,r)$
and provide a geometrical uplift of these theories via the compactification on spheres and hyperboloids.
To this end, the 56 internal coordinates are decomposed in the ${\rm SL}(8)$ frame
\bea
\{Y^M\} &=& \{Y^{[AB]}, Y_{[AB]}\}\;,\qquad
A,B=1, \dots, 8
\;,
\label{sl8}
\eea
into what we will refer to as `electric' and `magnetic' coordinates.
In \cite{Hohm:2014qga}, the physical coordinates are identified among the electric
$Y^{[AB]}$ as $y^i\equiv Y^{[i8]}$, corresponding to the $D=11$ solution  (\ref{coordinates})
of the section constraint.
In the ${\rm SL}(8)$ frame (\ref{sl8}), the latter takes the form
\bea
\partial_{AC} \otimes \partial^{BC}+\partial^{BC} \otimes \partial_{AC} &=&
\frac18\,\delta_A^B\left(\partial_{CD} \otimes \partial^{CD}
+\partial^{CD} \otimes \partial_{CD}\right)
\;,\nonumber\\
\partial_{[AB} \otimes \partial_{CD]} &=&
\frac1{24}\,
\varepsilon_{ABCDEFGH}\,\partial^{EF} \otimes \partial^{GH}
\;.
\label{secSL8}
\eea
The twist matrices $U_A{}^{{B}}(y^i)$ associated to sphere and hyperboloid compactifications
can then be constructed within the subgroup ${\rm SL}(8)\subset {\rm E}_{7(7)}$.

Here, we will generalize this result to twist matrices $U\subset {\rm SL}(8)$
which depend on more general subsets of coordinates (still satisfying the section constraint)
and take the form of products of the solutions found in \cite{Hohm:2014qga}. More precisely,
let us consider a twist matrix of the type
\bea
U(y^i,\tilde{y}_a)&\equiv& \mathring{U}(\tilde{y}_a)\,\hat{U}(y^i)\;,
\nonumber\\
\rho(y^i,\tilde{y}_a) &=& \mathring\rho(\tilde{y}_a)\,\hat\rho(y^i)
\;,
\label{UU}
\eea
where $\mathring{U}$ and $\hat{U}$ separately solve the Scherk-Schwarz consistency equations,
with embedding tensors denoted by $\mathring{X}_{MN}{}^K$ and $\hat{X}_{MN}{}^K$, respectively.
We also assume that $\hat\vartheta=\mathring\vartheta=0$\,.
With this ansatz, the first of the consistency equations (\ref{consistent_twist}) for $U$ reduce to
\bea
 \mathring\rho^{-1}\,\left[\Big(\hat{U}^{-1}\mathring{U}^{-1}\hat{U}\Big){}_{M}{}^N\,\hat{\cal X}_{NP}{}^Q\right]_{\bf 912}
+ \hat\rho^{-1}\; \hat{U}\!\left[\mathring{X}_{MP}{}^Q\right]
~\equiv~
{\rm const}
~\equiv~X_{MP}{}^Q
\;.
\label{UU1}
\eea
where $\hat{\cal X}_{MN}{}^K$ denotes the unprojected current
\bea
\hat{\cal X}_{MN}{}^K&\equiv&
\hat\rho^{-1}\,
(\hat U^{-1})_{{M}}{}^{P} (\hat U^{-1})_{{N}}{}^{Q} \, \partial_P \hat U_Q{}^{{K}}
\eea
(such that $[\hat{\cal X}_{MN}{}^K]_{\bf 912} = \hat{X}_{MN}{}^K$), and
\bea
\hat{U}\!\left[\mathring{X}_{MP}{}^Q\right] &\equiv&
(\hat{U}^{-1})_M{}^{M'} (\hat{U}^{-1})_N{}^{N'} \hat{U}_{K'}{}^K\,
\mathring{X}_{M'N'}{}^{K'}
\;,
\eea
denotes the ${\rm E}_{7(7)}$-action of $\hat{U}$ on the embedding tensor $\mathring{X}_{MP}{}^Q$\,.
Let us further assume that the variables $y^i$ and $\tilde{y}_a$ are mutually compatible in the sense that
\bea
\left(\mathring\rho^{-1}\,(\mathring{U}^{-1})_M{}^N\,\partial_N \right) \Big|_{y^i} &=& \partial_M \Big|_{y^i}
\;,
\nonumber\\
\left(\hat\rho^{-1}\,(\hat{U}^{-1})_M{}^N\,\partial_N \right) \Big|_{\tilde{y}_a} &=& \partial_M \Big|_{\tilde{y}_a}
\;,
\label{dero}
\eea
i.e.\ that we have equality of the action of these differential operators on the coordinates $y^i$ and $\tilde{y}_a$,
respectively.
With this assumption, the l.h.s.\ of equation (\ref{UU1}) reduces to
\bea
\hat{X}_{MP}{}^Q
+ \hat\rho^{-1}\; \hat{U}\!\left[\mathring{X}_{MP}{}^Q\right]
&=&
\hat{X}_{MP}{}^Q
+ \mathring{X}_{MP}{}^Q
\;,
\label{UU2}
\eea
such that equation (\ref{UU1}) is automatically satisfied with
the resulting embedding tensor given by
\bea
X_{MP}{}^Q ~=~
\hat{X}_{MP}{}^Q
+ \mathring{X}_{MP}{}^Q
\;.
\label{XXX}
\eea
We can introduce a relative coupling constant between $\hat{X}_{MP}{}^Q$ and $\mathring{X}_{MP}{}^Q$
by rescaling of the $\tilde{y}_a$ vs.\ the $y^i$ coordinates.
This allows us to capture the continuous deformation parameter of the $\SO(4)^2\ltimes T^{16}$ gaugings and of their non-compact forms.
Finally, the second equation of (\ref{consistent_twist}) turns into
 \bea
(\mathring{U}^{-1}){}_K{}^N \partial_N (\hat{U}^{-1})_M{}^K
+ (\hat{U}^{-1})_M{}^K \partial_N (\mathring{U}^{-1}){}_K{}^N
&=& 3\,\rho^{-1}\,(\hat{U}^{-1}\mathring{U}^{-1}){}_M{}^N\partial_N\rho
\;,
\eea
which together with (\ref{dero}) and the respective equations for $\hat\rho$ and $\mathring\rho$
turns into an identity.

In the following we will consider the product ansatz (\ref{UU})
with matrices $\hat{U}$ and $\mathring{U}^{-1}$ chosen among the solutions
from \cite{Hohm:2014qga}, corresponding to gauge groups ${\rm SO}(p,q)$
and ${\rm SO}(p',q')$, respectively.
In order to satisfy the compatibility constraints (\ref{dero}) together
with the section constraints (\ref{secSL8}), we will choose the coordinates $y^i$
among the electric and the $\tilde{y}_a$ among the magnetic coordinates from (\ref{sl8}).
More precisely, we define coordinates $\{y^i, \tilde{y}_a\}$
\bea
y^i \equiv Y^{i8}\;,\qquad \tilde{y}_a \equiv Y_{a7}
\;,\quad
i = 1, \dots,p+q-1\,;\quad
a=p+q, \dots, 6
\;,
\label{yia}
\eea
which provide a solution to the section constraints (\ref{secSL8}). Moreover, the associated
${\rm SL}(8)$ matrices $\hat{U}$ and $\mathring{U}$ commute, satisfy the compatibility equations (\ref{dero}),
and give rise to the product (\ref{UU})
\bea
({U}^{-1})_A{}^B &=&
\left(\mathring{\rho} \hat{\rho}^{-1}\right)^{1/2}
\left(
\begin{array}{c:c:c:c}
\hat{V}_i{}^j& 0& 0 & \hat{\rho}^{2}\,\hat{V}_i{}^0
\\ \hdashline
0 & \mathring{W}_a{}^b   & \mathring{\rho}^{-2}\,\mathring{W}_a{}^0  & 0
\\ \hdashline
0 & \mathring{\rho}^{-2}\,\mathring{W}_0{}^a & \mathring{\rho}^{-4}\,(1+\mathring{u} \mathring{K}(\mathring{u},\mathring{v}) )   & 0
\\ \hdashline
\hat{\rho}^{2}\,\hat{V}_0{}^j & 0 & 0 & \hat{\rho}^{4}
\end{array} \right)
\;,
\label{twistU}
\eea
which we present in the ${\rm SL}(8)$ basis $\{A\}\rightarrow\{i,a,7,8\}$\,.
The various blocks are given by
\bea
\label{Upq}
&&\hat{V}_0{}^i \equiv \eta_{ij} y^j\,\hat{K}(\hat{u},\hat{v})\;,
\quad
\hat{V}_i{}^0 \equiv \eta_{ij} y^j\,\;,
\quad
\hat{V}_i{}^j \equiv  \delta^{ij}\,+  \eta_{ik} \eta_{jl} \, y^k  y^l\,\hat{K}(\hat{u},\hat{v}) \;,
\nonumber\\[1ex]
&&\mathring{W}_0{}^a \equiv -\eta^{ab} \tilde{y}_b\,\;,
\qquad\
\mathring{W}_a{}^0 \equiv -\eta^{ab} \tilde{y}_b\,\mathring{K}(\mathring{u},\mathring{v})\;,
\quad
\mathring{W}_a{}^b \equiv  \delta^{ab} \;,
\eea
with $\eta_{ij}$ and $\eta^{ab}$ defining the signatures of ${\rm SO}(p-1,q)$ and
${\rm SO}(p'-1,q')$, respectively, and the functions $\hat\rho$, $\mathring\rho$ given by
\bea
\hat{\rho} &=& (1-\hat{v})^{1/4}~\equiv~(1-y^i\eta_{ij}y^j)^{1/4}\;,
\nonumber\\
\mathring{\rho} &=& (1-\mathring{v})^{1/4}~\equiv~(1-\tilde{y}_a\eta^{ab}\tilde{y}_b)^{1/4}\;.
\label{rhorho}
\eea
The functions $\hat{K}(\hat{u},\hat{v})$ and $\mathring{K}(\mathring{u},\mathring{v})$
are determined by first order differential equations and given explicitly in \cite{Hohm:2014qga}.
One may check explicitly that the matrix (\ref{twistU}) solves the consistency equations (\ref{consistent_twist})
and gives rise to the embedding tensor (\ref{embeddingX}) of the dyonic gaugings.
We stress that it is crucial for the consistency of the construction that the coordinates $y^i$ and $\tilde{y}_a$
are chosen within distinct ${\rm SL}(8)$ representations in (\ref{sl8}),
i.e.\ the $y^i$ and the $\tilde{y}_a$ are embedded in the electric and magnetic coordinates, respectively.

\subsection{Type II origin}\label{s:5}

In the previous section, we have constructed the Scherk-Schwarz twist matrices that give
rise to the embedding tensor of dyonic gaugings. Since we have identified the coordinates $\{y^i, \tilde{y}_a\}$
on which these matrices depend directly in the ${\rm SL}(8)$ frame (\ref{sl8}), it is not immediately obvious
if these coordinates in the ${\rm GL}(6)$ bases (\ref{coordinates}) correspond to a IIA or IIB solution
of the section constraints. We will determine their precise higher-dimensional origin case by case
according to the value of $p+q$.

\paragraph{{\em p}+{\em q}\,=\,6 :}

In this case, the coordinates (\ref{yia}) are given by $\{Y^{18}, Y^{28}, Y^{38}, Y^{48}, Y^{58}, Y_{67}\}$.
Comparing this set to the section constraint (\ref{secSL8}), it follows that fields can depend on none of the other
50 internal coordinates without violating the section constraint.
We conclude that exceptional field theory on this set of coordinates is equivalent to IB supergravity.
More specifically, we can identify the ${\rm SL}(2)_{\rm IIB}$ under which these coordinates are singlets as the
subgroup of ${\rm SL}(8)$ whose generators are given by
\bea
{\rm SL}(2)_{\rm IIB} &=& \left\langle T_6{}^7, T_7{}^6,T_7{}^7-T_6{}^6 \right\rangle
\;,
\label{SL26}
\eea
where E$_{7(7)}$ generators are defined in appendix~\ref{app:generators}.
The ${\rm GL}(1)_{\rm IIB}\subset{\rm G}_{\rm IIB}$ which provides the geometric grading of
coordinates (\ref{coordinates}) and fields is generated by
\bea
{\rm GL}(1)_{\rm IIB} &=& \left\langle T_8{}^8 -\frac12 \left( T_6{}^6+T_7{}^7\right) \right\rangle
\;.
\label{GLpq6}
\eea
Indeed, evaluating the charges of the various coordinates under this ${\rm GL}(1)_{\rm IIB}$, we find
\bea
\{Y^{i8},Y_{67}\}: -4\;,\quad
\{Y^{a8},Y_{ia}\} :  -2 \;,\quad
\{Y^{ij}, Y_{ij}\}  :   0 \;,\quad \dots
\;,
\label{coord6}
\eea
thus reproducing the IIB charges of (\ref{coordinates}).

\paragraph{{\em p}+{\em q}\,=\,5 :}

In this case, the coordinates (\ref{yia}) are given by $\{Y^{18}, Y^{28}, Y^{38}, Y^{48}, Y_{57}, Y_{67}\}$.
It is straightforward to verify that they can be extended by a seventh coordinate $Y_{56}$ still
satisfying the section constraints (\ref{sl8}). The resulting theory thus is Type IIA supergravity (with possible $D=11$ embedding).
The ${\rm GL}(1)_{\rm IIA}$ which provides the geometric grading of
coordinates (\ref{coordinates}) and fields is generated by
\bea
{\rm GL}(1)_{\rm IIA} &=& \left\langle \frac34\left(T_8{}^8 -T_7{}^7\right) - \frac12 \left( T_5{}^5+ T_6{}^6\right) \right\rangle
\;,
\eea
giving rise to the charges
\bea
\{Y^{i8}, Y_{a7}\}: -4\,,\quad
\{Y^{ab}\}  :  -3 \,,\quad
\{Y^{a8}, Y_{i7}\}: - 2 \,,\quad
\{Y^{ij}, Y^{78}, Y_{ia}\}  :   -1 \,\quad \dots\;,
\eea
for the coordinates, in accordance with (\ref{coordinates}).

\paragraph{{\em p}+{\em q}\,=\,4 :}

In this case, the coordinates (\ref{yia}) are given by $\{Y^{18}, Y^{28}, Y^{38}, Y_{47}, Y_{57}, Y_{67}\}$.
As for $p+q=6$, it is straightforward to see that these coordinates cannot be extended by any of the other
50 internal coordinates without violating the section constraint.
Again, the resulting theory thus is IIB.
The ${\rm SL}(2)_{\rm IIB}$ under which these coordinates are singlets is not
entirely contained in ${\rm SL}(8)$ but has generators given by
\bea
{\rm SL}(2)_{\rm IIB} &=& \left\langle \left(T_8{}^8+T_4{}^4+T_5{}^5+ T_6{}^6\right), T_{4568}, T_{1237}
 \right\rangle
\;,
\eea
the latter two of which sit in the ${\bf 70} = \mathfrak{e}_{7(7)}\backslash\, \mathfrak{sl}(8)$\,.
The ${\rm GL}(1)_{\rm IIB}$ which provides the geometric grading of
coordinates (\ref{coordinates}) and fields is generated by
\bea
{\rm GL}(1)_{\rm IIB} &=& \left\langle \frac34 \left(T_8{}^8 -T_7{}^7\right) +
\frac14 \left( T_1{}^1+T_2{}^2+T_3{}^3-T_4{}^4-T_5{}^5- T_6{}^6\right) \right\rangle
\;,
\eea
giving charges
\bea
\{Y^{i8}, Y^{a7}\} : -4\,,\quad
\{Y^{a8}, Y^{ij}, Y_{i7}, Y_{ab}\}  :  -2 \,,\quad
\{ Y^{ia}, Y^{78}, Y_{78}, Y_{ia} \} : 0 \;,\quad
\dots\;,
\eea
for the coordinates, in accordance with (\ref{coordinates}).

\paragraph{{\em p}+{\em q}\,=\,3 :}

In this case, the coordinates are given by $\{Y^{18}, Y^{28}, Y_{37}, Y_{47}, Y_{57}, Y_{67}\}$.
Upon flipping $Y^{AB}\leftrightarrow Y_{AB}$, this choice maps into the case of $p+q=5$ above,
it thus corresponds to a IIA embedding of the theory.

\paragraph{{\em p}+{\em q}\,=\,2 :}

In this case, the coordinates are given by $\{Y^{18}, Y_{27}, Y_{37}, Y_{47}, Y_{57}, Y_{67}\}$.
Upon flipping $Y^{AB}\leftrightarrow Y_{AB}$, this choice maps into the case of $p+q=6$ above,
it thus corresponds to a IIB embedding of the theory.

\section{Uplift of the ${\cal N}=4$ AdS$_4$ vacuum}\label{s:6}

\subsection{${\cal N}=4$ AdS$_4$ vacuum in $D=4$ supergravity}

The $(\rm SO(6)\times SO(1,1))\ltimes T^{12}$ gauged maximal supergravity admits an $\mathcal N=4$ AdS$_4$ vacuum preserving SO(4) gauge symmetry \cite{Gallerati:2014xra}.
This solution is part of a one-parameter family of ${\cal N}=4$ AdS vacua belonging to inequivalent gauged maximal supergravities but exhibiting similar physical properties.
The other elements of this family of solutions are vacua of the $\omega$-deformed SO(7,1) gauged supergravities, whose higher-dimensional origin is unknown.
However, at a singular point in the parameter space of the family the gauging degenerates into $(\rm SO(6)\times SO(1,1))\ltimes T^{12}$, for which we can now provide an uplift to Type IIB supergravity.

Using \eqref{twistU} for $p=6,\,q=0$ and $p'=q'=1$, the Scherk--Schwarz ansatz \eqref{SSansatz} describes the consistent truncation of Type IIB supergravity to $(\rm SO(6)\times SO(1,1))\ltimes T^{12}$ gauged maximal $D=4$ supergravity described by the embedding tensor \eqref{embeddingX} with
\begin{align}
\eta_{AB} &= \mathrm{diag}(-1,\ldots,-1,0,0,-1)\,,\quad
\tilde\eta^{AB} = \mathrm{diag}(0,\ldots,0,-1,1,0)\,.
\label{theta xi SO6SO11}
\end{align}

In order to uplift the $\mathcal N=4$ AdS$_4$ solution of \cite{Gallerati:2014xra} we will need to reproduce the vacuum extremizing the scalar potential \eqref{sugra scal pot} in terms of the scalar matrix
\bea
M_{{MN}} = (L \,L^T)_{{MN}}&=&
\begin{pmatrix}
M_{AB,CD} & M_{AB}{}^{CD}\\
M^{AB}{}_{CD} & M^{AB,CD}
\end{pmatrix}
\;,
\label{MSL8}
\eea
expanded in the ${\rm SL}(8)$ basis (\ref{sl8}).
Here $L$ is a coset representative for $\rm E_{7(7)}/SU(8)$.
The $\mathcal N=4$ AdS$_4$ vacuum is located in an SO$(4)\subset\mathrm{SO(6)}$ invariant subspace of the scalar manifold, which turns out to be a GL(3)/SO(3) sub-coset space generated by \cite{DallAgata:2012sx}
\begin{align}
t_1 &=\tfrac1{12\sqrt2}( T_1{}^1+T_2{}^2+T_3{}^3-T_4{}^4-T_5{}^5-T_8{}^8)\,,\\
t_2 &=\tfrac1{24\sqrt2}( T_1{}^1+T_2{}^2+T_3{}^3+T_4{}^4+T_5{}^5+T_8{}^8-3T_6{}^6-3T_7{}^7) \,,\\
t_3 &= \tfrac1{4\sqrt6}( T_6{}^6-T_7{}^7 )\,,\\
t_4 &= \tfrac1{4\sqrt6}(T_6{}^7+T_7{}^6) \,,\ \
t_5 = \tfrac{1}{4\sqrt3}T_{1236} \,,\ \
t_6 = -\tfrac{1}{4\sqrt3}T_{1237} \,.
\end{align}
We normalize these generators so that $\mathrm{Tr}_{\bf56}(t_i t_i^T) = 1$ in the fundamental of E$_{7(7)}$.

Actually only some of the fields associated with these generators acquire a non-trivial value at the $\mathcal N=4$ AdS$_4$ vacuum.
We find that the vacuum solution is identified with the coset representative
\begin{align}
 L &= \exp\left(-\tfrac{3}{\sqrt2}\log(3) \, t_1 \right) \, \exp\left( \pm4\sqrt3 \, t_5 \pm4\sqrt3 \, t_6  \right)\ .
\label{vacuum coset}
\end{align}
Upon computation of the fermion shifts associated with this extremum of the scalar potential, we explicitly recover the solution of \cite{Gallerati:2014xra} up to an SU(8) transformation.
The $\pm$ signs in the coset representative give different instances of equivalent vacua.
We will take both negative in the following.
We give the explicit form of the scalar matrix $M_{ MN}$ at the vacuum in appendix~\ref{app:vacuumM}.

There are of course flat directions of the solution \eqref{vacuum coset} associated with the broken gauge symmetries.
The flat direction associated with the broken SO(1,1) will be relevant in the following.
It corresponds to mapping \eqref{vacuum coset} into the gauge-equivalent solution
\be
L \to \exp\left(\xi\,t_4\right) \, L \,.
\label{flat SO11}
\ee
Moreover, there are different instances of this vacuum connected by discrete transformations.
Beyond the signs indicated above, also the outer automorphism of the residual SO(4) that exchanges its SO(3) factors generates new instances of this vacuum in field space.
These are obtained by the following substitutions in \eqref{vacuum coset}:
\begin{equation}
t_1\to-t_1\,,\quad
t_{5,6} \to t_{5,6}^T \,.
\end{equation}

\subsection{Uplift formulas from generalized Scherk-Schwarz reduction}

The explicit uplift formulas that provide the embedding of the four-dimensional gauging into the IIB
theory are straightforwardly obtained by combining the Scherk-Schwarz ansatz (\ref{SSansatz}) with the dictionary
between the IIB theory and ${\rm E}_{7(7)}$ ExFT under the corresponding solution of the section constraint.
Here we are interested in the uplift of a special class of four-dimensional solutions, that preserve the AdS$_4$
isometries. In the four-dimensional theory, other than the external AdS$_4$ metric,
only scalar fields are excited and take constant values. Accordingly, among the ExFT fields (\ref{fcontent}) only
external and internal metric $g_{\mu\nu}$, ${\cal M}_{MN}$ are non-vanishing. The match of the latter with the IIB fields
is found upon breaking ${\rm E}_{6(6)}$ under the ${\rm GL}(6)\times {\rm SL}(2)$ that defines the IIB coordinates
(\ref{coordinates}). Explicitly, we denote the decomposition of the 56 internal coordinates as
\bea
\{Y^M\} &\longrightarrow&
\{
\tilde Y^m, \tilde Y_{m\alpha}, \tilde Y^{kmn}, \tilde Y^{m\alpha}, \tilde Y_m
\}\;,
\label{coordGL6}
\eea
with $m=1, \dots, 6$ and $\alpha=1, 2$ labeling the fundamental representations of ${\rm SL}(6)$ and
${\rm SL}(2)$, respectively. Accordingly, the matrix ${\cal M}_{MN}$ decomposes into blocks
\bea
{\cal M}_{MN} &=&
{\footnotesize
\begin{pmatrix}
{\cal M}_{m,m'} & {\cal M}_m{}^{m'\beta} & {\cal M}_{m,m'n'p'} & {\cal M}_{m,m'\beta} & {\cal M}_{m}{}^{m'}
\\
{\cal M}^{m\alpha}{}_{m'} & {\cal M}^{m\alpha,m'\beta} & {\cal M}^{m\alpha}{}_{m'n'p'} & {\cal M}^{m\alpha}{}_{m'\beta} & {\cal M}^{m\alpha,m'}
\\
{\cal M}_{mnp,m'} & {\cal M}_{mnp}{}^{m'\beta} & {\cal M}_{mnp,m'n'p'} & {\cal M}_{mnp,m'\beta} & {\cal M}_{mnp}{}^{m'}
\\
{\cal M}_{m\alpha,m'} & {\cal M}_{m\alpha}{}^{m'\beta} & {\cal M}_{m\alpha,m'n'p'} & {\cal M}_{m\alpha,m'\beta} & {\cal M}_{m\alpha}{}^{m'}
\\
{\cal M}^{m}{}_{m'} & {\cal M}^{m,m'\beta} & {\cal M}^{m}{}_{m'n'p'} & {\cal M}^{m}{}_{m'\beta} & {\cal M}^{m,m'}
\end{pmatrix}
}
\;.
\label{blockM}
\eea
The explicit form of these blocks is read off from expanding the exponential series (\ref{V56}) and
(after proper normalization) gives rise to the following identification of the IIB fields
\bea
{\cal M}^{mn} &=& G^{-1/2}\,G^{mn} \;,
\nonumber\\
{\cal M}^{m}{}_{n\alpha} &=&  \frac{1}{\sqrt{2}}\,G^{-1/2}\,G^{mk}\,b_{kn}{}^{\beta}\,\varepsilon_{\beta\alpha}\;,
\nonumber\\
{\cal M}_{m\alpha,n\beta} &=&
\frac12\,G^{-1/2}\,G_{mn}\,m_{\alpha\beta} + \frac12\,G^{-1/2}\,G^{kl}\, b_{mk}{}^\gamma b_{nl}{}^\delta\,\varepsilon_{\alpha\gamma}\varepsilon_{\beta\delta}
\;,\nonumber\\
{\cal M}^p{}_{lmn} &=&  -2\,G^{-1/2}\,G^{pk} \left(c_{klmn} -\frac38\, \varepsilon_{\alpha\beta}\, b_{k[l}{}^\alpha b_{mn]}{}^\beta\right)
\;.
\label{compM}
\eea
For the uplift formulas we need to evaluate the l.h.s.\ of these expressions
via the Scherk-Schwarz ansatz (\ref{SSansatz})
\bea
{\cal M}_{MN}(x,Y) &=& U_M{}^{{M}}(y) U_N{}^{{N}}(y)\, M_{{MN}}(x)
\;,\label{sMM}
\eea
with the ${\rm SL}(8)$ valued twist matrix $U$ from (\ref{twistU}). In order to reconcile the
${\rm GL}(6)\times {\rm SL}(2)$ decomposition of (\ref{blockM}) with the ${\rm SL}(8)$ form of
the twist matrix, we have to break both groups down to their common ${\rm SL}(5)\times {\rm SL}(2)$.
For the coordinates (\ref{coordGL6}) this implies
\bea
\{Y^M\} &\longrightarrow&
\{
\tilde Y^m, \tilde Y_{m\alpha}, \tilde Y^{kmn}, \tilde Y^{m\alpha}, \tilde Y_m
\} \nonumber\\
&\longrightarrow&
\{\tilde Y^i, \tilde Y^6, \tilde Y_{i\alpha}, \tilde Y_{6\alpha}, \tilde Y^{ijk} , \tilde Y^{6ij}, \tilde Y^{i\alpha}, \tilde Y^{6\alpha}, \tilde Y_i, \tilde Y_6
\}
\;,
\label{YYY}
\eea
upon splitting $\{m\}\rightarrow\{i,6\}$\,. Similarly, for the ${\rm SL}(8)$ coordinates (\ref{sl8}),
and in accordance with (\ref{coord6}), we use the split of ${\rm SL}(8)$ indices
\bea
\{A\} &\longrightarrow& \{\underline{i},\underline{a}\}\;,
\quad\mbox{with}
\quad
\underline{i} = \{i,8\}\;,\quad i=\{1,\dots, 5\}\;,\quad
\underline{a}=\{6,7\}
\;,
\label{splitia}
\eea
in order to decompose the $\{Y^{[AB]}, Y_{[AB]}\}$. We may then identify the coordinates (\ref{YYY}) among
the ${\rm SL}(8)$ coordinates (\ref{sl8}) as
\bea
&&{}
\{\tilde Y^i, \tilde Y^6, \tilde Y_{i\alpha}, \tilde Y_{6\alpha}, \tilde Y^{ijk} , \tilde Y^{6ij}, \tilde Y^{i\alpha}, \tilde Y^{6\alpha}, \tilde Y_i, \tilde Y_6
\}
\nonumber\\
&&{}\qquad\qquad\qquad
~=~
\{Y^{i8}, Y_{67}, Y_{i\underline{a}}, \varepsilon_{\underline{ab}}Y^{\underline{b}8} , \varepsilon^{ijki'j'} Y_{i'j'}, Y^{ij}, Y^{i\underline{a}}, \varepsilon^{\underline{ab}} Y_{\underline{b}8} , Y_{i8}, Y^{67}
\}\;,
\eea
where according to (\ref{SL26}) we identify $\{\underline{a}\}=\{6,7\}$ from (\ref{splitia})
with the ${\rm SL}(2)$ doublet indices $\{\alpha\}=\{1,2\}$\,.

Let us now make the uplift formulas explicit.
Combining (\ref{compM}) with (\ref{sMM})
and the form of the twist matrix (\ref{twistU}), we obtain
\bea
 G^{-1/2}\,G^{ij} &=& {\cal M}^{ij} ~=~ 2\,{\cal M}^{i8,j8}
 \nonumber\\
 &=& 2\,(U^{-1})_{\underline{kl}}{}^{i8}(U^{-1})_{\underline{mn}}{}^{j8}\,M^{\underline{kl},\underline{mn}}(x)
 \nonumber\\
 &=&
 2\, \rho^2 \, {\cal K}_{\underline{kl}}{}^{i}{\cal K}_{\underline{mn}}{}^{j}\,M^{\underline{kl},\underline{mn}}(x)
 \;,
\eea
where $M^{\underline{kl},\underline{mn}}(x)$ refers to part of the lower right block of the ${\rm E}_{7(7)}$ matrix
(\ref{MSL8}), and we have expressed the relevant components of the twist matrix $U$ in terms
of the Killing vectors on the round five-sphere
\bea
{\cal K}_{\underline{mn}}{}^i ~=~ \hat{G}^{ij}\,\partial_j {\cal Y}_{[\underline{m}}\, {\cal Y}_{\underline{n}]}
\;,\qquad
{\cal Y}_{\underline{m}} &\equiv& \{ y^i, \sqrt{1-y^iy^i}\}
\;,
\nonumber\\
\hat{G}^{ij} &\equiv& \delta^{ij} - y^i y^j
\;.
\label{KV}
\eea
Similar calculation determines the remaining components of the internal six-dimensional metric, such that together we find
\bea
G^{ij} &=&
2\, \Delta \, {\cal K}_{\underline{kl}}{}^{i}{\cal K}_{\underline{mn}}{}^{j}\,M^{\underline{kl},\underline{mn}}(x)
\;,\nonumber\\
 G^{i6} &=&
 2\,\Delta \,  \mathring{\rho}^2\, {\cal K}_{\underline{kl}}{}^{i}M^{\underline{kl}}{}_{\underline{67}}(x)
 \;,\nonumber\\
 G^{66} &=&
 2\,\Delta \,\mathring{\rho}^4\,M_{\underline{67},\underline{67}}(x)
 \;,
 \label{Glift}
\eea
with $\mathring{\rho}$ from (\ref{rhorho}) and the scale factor $\Delta$ defined by
\bea
\Delta&\equiv& \rho^2\,({\rm det}\,G)^{1/2}
\;.
\label{Delta}
\eea
While (\ref{Glift}) represent the uplift formulas for generic solutions of the four-dimensional theory,
in the vacuum (\ref{vacuum coset}) we are interested in lifting, the matrix $M_{MN}(x)$ is constant, and
these formulas further reduce to
\bea
G^{ij} &=&
\left\{
\begin{array}{rcl}
(1+2r^2)\,\Delta\,\delta^{ij}-3 \,\Delta\,y^iy^j&:& i,j\in\{1,2,3\}\\
(3-2r^2)\,\Delta\,\delta^{ij}-3 \,\Delta\,y^iy^j &:& i,j\in\{4,5,6\}\\
-\Delta\,y^iy^j&:& i\in\{1,2,3\}\,,\;j\in\{4,5,6\}\\
\end{array}
\right.
\;,\nonumber\\[1ex]
 G^{i6} &=&
 0\;,\qquad
 G^{66} ~=~
(1-\tilde{y}_6^2)\, \Delta
 \;,
 \label{Glift0}
\eea
with $r^2\equiv (y^1)^2+(y^2)^2+(y^3)^2 \le1$  and
\bea
\Delta &=& \left((1+2r^2)(3-2r^2)\right)^{-1/4}
\;.
\label{Delta0}
\eea

In a similar way, we may obtain the uplift formulas for the remaining IIB fields from (\ref{compM}).
For the two-form, we find that its only non-vanishing components are given by
\bea
b_{ij}{}^{\alpha}&=&
2\,G^{1/2}\,G_{ik}\,\varepsilon^{\underline{ab}}\,{\cal M}^{k8}{}_{j\underline{b}}  ~=~
4\,G^{1/2}\,G_{ik}\,\varepsilon^{\underline{ab}}(U^{-1})_{\underline{kl}}{}^{k8}\, U_{j\underline{b}}{}^{\underline{mc}} \,
M^{\underline{kl}}{}_{\underline{mc}}
\nonumber\\
&=&
-2\,\Delta\,G_{ik}\,{\cal K}_{\underline{kl}}{}^{k}
\partial_j{\cal Y}^{\underline{m}}
\varepsilon^{\underline{cd}} A_{\underline{d}}{}^{\alpha}
\,
M^{\underline{kl}}{}_{\underline{mc}}\,
\;,
\label{b0}
\eea
where as above we identify $\{\underline{a}\}=\{6,7\}$ from (\ref{splitia})
with the ${\rm SL}(2)$ doublet indices $\{\alpha\}=\{1,2\}$\,,
and $M^{\underline{kl}}{}_{\underline{mc}}$ is given in (\ref{a2}).
The ${\rm SL}(2)$ matrix $A_{\underline{a}}{}^{\underline{b}}$ is read off as
\bea
A_{\underline{a}}{}^{\alpha}&\equiv&
\begin{pmatrix}
\mathring\rho^{2} & \tilde{y}_6\\
 \tilde{y}_6 & \mathring\rho^{-2}(1+\tilde{y}_6^2)
\end{pmatrix}
~=~
\begin{pmatrix}
\sqrt{1+\tilde{y}_6^2} & \tilde{y}_6\\
 \tilde{y}_6 & \sqrt{1+\tilde{y}_6^2}
\end{pmatrix}
 \;,
\label{matA}
\eea
from the $(6,7)$ block of (\ref{twistU}), using that $\mathring{K}=1$ in this case.

Next, the IIB dilaton/axion matrix is obtained from (\ref{compM}) as
\bea
m_{\alpha\beta}&=&
\frac13\,G \left({\cal M}^{mn} \, {\cal M}_{m\alpha,n\beta}
-4\,{\cal M}^n{}_{k\alpha} {\cal M}^k{}_{n\beta} \right)
\;,
\eea
which when put together with (\ref{sMM}) in our vacuum yields
\bea
m_{\alpha\beta} &=&
\frac23\,\Delta^2 \,{\cal Y}_{\underline{m}}\,{\cal Y}_{\underline{n}}\,
S^{\underline{ma},\underline{nb}}
\,
A_{\underline{a}}{}^{(\alpha'}A_{\underline{b}}{}^{\beta')}
\varepsilon_{\alpha\alpha'}\varepsilon_{\beta\beta'}
\;,
\eea
with the matrix $A_{\underline{a}}{}^\alpha$ from (\ref{matA}) and
\bea
S^{\underline{ma},\underline{nb}} &=&
\frac12\,M^{\underline{ma},\underline{nb}}
+
\varepsilon^{\underline{ac}}
\varepsilon^{\underline{bd}}\left(
\,M_{\underline{kc},\underline{ld}}\,M^{\underline{km},\underline{ln}}
+
M^{\underline{km}}{}_{\underline{lc}}
M^{\underline{ln}}{}_{\underline{kd}}
\right)
\;.
\eea
With the explicit values (\ref{a1})--(\ref{a6}) of $M_{MN}$ in our vacuum this expression reduces to
\bea
m_{\alpha\beta} &=& (A^{-1} M A^{-T})_{\alpha\beta}\;,
\qquad
M_{\underline{ab}} ~\equiv~ \frac{\Delta^2}{\sqrt{3}}\,
\begin{pmatrix}
3+2r^2&-4r^2\\
-4r^2&3+2r^2
\end{pmatrix}
\;,
\label{mmm}
\eea
and it comes as a non-trivial consistency check, that with the expression (\ref{Delta0})
for the scale factor $\Delta$, this matrix indeed has determinant 1.

Finally, the expression for the only non-vanishing components of the
IIB 4-form follows from
\bea
{\cal M}^i{}_{jkl} &=&
\frac12\,{\cal M}^{i8,j'k'} \varepsilon_{jklj'k'}
~=~\frac12\,\rho\,\varepsilon_{jklj'k'}\,
{\cal K}_{\underline{kl}}{}^{i}(U^{-1})_{\underline{mn}}{}^{j'k'}
{M}^{\underline{kl},\underline{mn}}
\nonumber\\
&=&
\frac1{2\Delta} \, \rho^2\,
G^{ii'} \,\hat{C}_{jkli'}
-
\frac12\,\rho^2\,\hat\rho^{-2}\, \varepsilon_{jklj'k'}\,
{\cal K}_{\underline{kl}}{}^{i} \hat\nabla^{j'} {\cal K}_{{\underline{mn}}}{}^{k'}
{M}^{\underline{kl},\underline{mn}}
\;,
\eea
with $\hat{C}_{ijkl}$ defined as giving rise to the $S^5$ background flux
\bea
5\,\partial_{[i'} \hat{C}_{ijkl]} &=& \hat{\omega}_{i'ijkl}~\equiv~\hat\rho^{-2}\,\varepsilon_{i'ijkl}
\;.
\eea
Together, the expression for the IIB 4-form is given as
\bea
c_{ijkl} &=&
\hat{C}_{ijkl}
+
\frac14\, \Delta\,
{\cal K}_{\underline{kl}}{}^m G_{m[i}  \hat\omega_{jkl]j'k'}\, \hat\nabla^{j'} {\cal K}_{{\underline{mn}}}{}^{k'}
{M}^{\underline{kl},\underline{mn}}
\;.
\label{c0}
\eea
We have thus obtained all the non-vanishing IIB fields as functions of the $S^5$ Killing vectors and
sphere harmonics. Let us note that the expansion (\ref{V56}) also carries some components
$b_\alpha\equiv \epsilon^{klmnpq}\,b_{klmnpq\,\alpha}$ of the dual six-form of the IIB theory
which however vanish identically in our vacuum.

\subsection{The supersymmetric IIB AdS$_4 \times M_5\times S^1$ solution}
\label{sec:solution}

In this section, we calculate the field strengths and present the IIB solution in its most compact form.
The vacuum (\ref{vacuum coset}) of the four-dimensional theory preserves ${\cal N}=4$ supersymmetry and
accordingly a global ${\rm SO}(4)={\rm SO}(3)\times {\rm SO}(3)$ symmetry that shows up as the internal isometry group of
the IIB solution. In order to make these isometries manifest, we split the
$S^5$ sphere harmonics into
\bea
\left\{{\cal Y}^{\underline{m}} \right\} &=&
\left\{ {\cal Y}^p,\,{\cal Z}^{p}\equiv{\cal Y}^{p+3}\right\}\;,\quad p=1,2,3\;,
\quad {\cal Y}^p{\cal Y}^p=1-{\cal Z}^{p}{\cal Z}^{p}\equiv r^2
\;.
\eea
In terms of these harmonics, the ten-dimensional IIB metric is given by
\bea
ds^2 &=&
\Delta^3 \left( (3-2r^2)\,\delta^{pq}+ 8\, {\cal Y}^p {\cal Y}^q\right) d{\cal Y}^pd{\cal Y}^q  +
\Delta^3 \left(1+2r^2\right)  d{\cal Z}^p d{\cal Z}^p
\nonumber\\[1ex]
&&{}+\Delta^{-1}\left(d\eta\, d\eta + \frac{1}{2}\,ds^2_{{\rm AdS_4}}
\right)
\;,\label{IIBmetric}
\eea
with the warp factor given by (\ref{Delta0}) as
\bea
\Delta &=& \left((1+2r^2)(3-2r^2)\right)^{-1/4}
\;,
\eea
and the AdS$_4$ radius fixed to $r_{\rm AdS}=1$\,.
W.r.t.\ the previous sections, we have also changed coordinates $\tilde{y}_6={\rm sinh}\,\eta$
along the $S^1$ direction.
The internal five-dimensional space is a deformation of the round metric on $S^5$ which preserves an ${\rm SO}(3)\times{\rm SO}(3)\subset {\rm SO}(6)$ of the isometry group.
Indeed, the harmonics $\mathcal Y^p,\,\mathcal Z^p$ can be regarded as embedding coordinates for two $S^2$ spheres of radii $r$ and $\sqrt{1-r^2}$, respectively. 
The $S^5$ geometry is parametrized in terms of these two spheres fibered over the interval $r\in(0,1)$, and at the points $r=0,1$ one of the $S^2$'s shrinks smoothly to zero size.
Denoting $d\Omega_{1,2}^2$ the round metrics of unit radius on the $S^2$'s, an explicit expression for \eqref{IIBmetric} is
\bea
\nonumber
ds^2 &=&
\Delta^3 (3-2r^2)\, r^2 d\Omega_1^2 + \Delta^3(1+2r^2)\,(1-r^2) d\Omega_2^2\\[1ex]
&&+\Delta^{-1}\left( d\eta^2+\frac{dr^2}{1-r^2} +\frac12\,ds^2_{{\rm AdS_4}} \right)
\;.
\eea

The ${\rm SL}(2)$ matrix of IIB supergravity
\bea
m_{\alpha\beta}&=&
\frac{1}{\Im\, \tau}\begin{pmatrix}
|\tau|^2&-\Re\,\tau\\
-\Re\,\tau&1
\end{pmatrix}
\;,
\qquad
\tau~=~ C_0+i e^{-\Phi}
\;,
\label{standard axiodil param}
\eea
describing dilaton and axion is given by (\ref{mmm}) as
\bea
m_{\alpha\beta} &=& (A^{-1} M A^{-T})_{\alpha\beta}\;,
\eea
as a product of the ${\rm SL}(2)$ matrices
\bea
M_{\alpha\beta} &\equiv& \frac{\Delta^2}{\sqrt{3}}\,
\begin{pmatrix}
3+2r^2&-4r^2\\
-4r^2&3+2r^2
\end{pmatrix}
\;,
\quad
A_{\alpha}{}^{\beta}~\equiv~
\begin{pmatrix}
{\rm cosh}\,\eta & {\rm sinh}\,\eta\\
 {\rm sinh}\,\eta & {\rm cosh}\,\eta
\end{pmatrix}
\;.
\label{AM}
\eea
The three-form field strength is obtained by exterior derivative of (\ref{b0}) and takes the form
\bea
H_3{}^\alpha &=&
\frac{(V_- A){}^\alpha}{1+2r^2} \,\varepsilon_{pqr}
\,d{\cal Y}^p\wedge d{\cal Y}^q\wedge
\left(
\frac{(3+2r^2)}{3\,(1+2r^2)} \,
d{\cal Y}^r-{\cal Y}^r\,  d\eta\right)
\nonumber\\
&&{}
- \frac{ (V_+ A){}^\alpha}{3-2r^2}
\,\varepsilon_{pqr}\,d{\cal Z}^p\wedge d{\cal Z}^q\wedge
\left(
\frac{(5-2r^2)}{3\,(3-2r^2)}\,
d{\cal Z}^r + {\cal Z}^r\,  d\eta\right)
\;,
\eea
with the matrix $A$ from (\ref{AM}) and the vectors
\bea
V_\pm^{\alpha}=\{3^{\pm1/4},\pm3^{\pm1/4}\}\;.
\eea
Finally, the selfdual IIB five-form field strength is given by
\bea
H_5 &=& d c -\frac18\,\varepsilon_{\alpha\beta}\,b^\alpha \wedge db^\beta
\nonumber\\
&=&
\frac{6\Delta^4}{8\,(1-r^2)}
{\cal Z}^{p}\,d{\cal Y}^p \wedge d{\cal Y}^q\wedge d{\cal Y}^r\wedge  d{\cal Z}^q\wedge d{\cal Z}^r
\nonumber\\
&&{}
+3\,\Delta^4
\,{\cal Z}^{p} {\cal Y}^{[p} \,d{\cal Y}^q\wedge d{\cal Y}^{r]}\wedge d{\cal Z}^q\wedge d{\cal Z}^r\wedge d\eta
\nonumber\\
&&{}
-\frac1{16}\,\sqrt{|g|}\,\varepsilon_{\mu\nu\rho\sigma} \, dx^\mu \wedge dx^\nu \wedge dx^\rho \wedge dx^\sigma \wedge
\left(
d\eta  -\frac43\, {\cal Y}^p d{\cal Y}^p
\right)
\;.
\eea
We have explicitly verified, that this solution satisfies all the field equations of the IIB theory,
including the Einstein equations.

\subsection{Interpretation, S-folds and supersymmetry}\label{sfoldsuper}

The uplift provided in the previous section is in principle on a warped $S^5\times\mathbb R$ internal space, where $\mathbb R$ is the $\eta$ direction.
In fact, $\partial/\partial\eta$ is an isometry of the solution and although the $\eta$ dependence is present in the fluxes, it only appears through the SL(2) matrix $A(\eta)$ of \eqref{AM}.
Indeed, the flat direction \eqref{flat SO11} lifts to constant shifts of $\eta$.
This means that we can make $\eta$ periodic, $\eta\simeq\eta+T$, at the price of introducing an SL(2) monodromy of the fields along the resulting $S^1$:
\begin{equation}
\mathfrak M_{S^1} = A(\eta)^{-1} A(\eta+T) \,.
\label{monodromy}
\end{equation}
Being $A(\eta)$ an element of a non-compact subgroup of SL(2), there is no choice of the period $T$ such that $\mathfrak M_{S^1} =1$, which would make the solution globally geometric.
Instead, the solution is locally geometric and globally an S-fold.

The periodicity in $\eta$ is restricted if we require that the resulting monodromy belongs to SL$(2,\mathbb Z)$.
For instance, to obtain the representatives of the infinite sequence of hyperbolic SL$(2,\mathbb Z)$ conjugacy classes (see e.g.\ \cite{Dabholkar:2002sy})
\begin{equation}
\label{hyperbolic conj classes}
\mathfrak M(n) =
\begin{pmatrix}
n & 1\\ -1 & 0
\end{pmatrix},\quad \ n\in\mathbb N\,,\ n \ge 3\,,
\end{equation}
we must set $T=\log\tfrac12(n + \sqrt{n^2-4})$ and redefine $A(\eta)$ in all expressions (including the Scherk--Schwarz matrix \eqref{twistU}) as follows:\footnote{Note that $g$ is not unique, it can be redefined by $g \to \exp(\zeta\log\mathfrak{M}_{S^1}) g$ for any $\zeta$.
Also note that conjugacy classes with $n<-3$ cannot be obtained from our initial monodromy matrix.}
\begin{equation}
A(\eta) \to A(\eta) g\,,\qquad
g\equiv
\left(\begin{matrix}
\frac{(n^2-4)^{1/4}}{\sqrt2}\ & 0\\
\frac{n}{\sqrt2 (n^2-4)^{1/4}}
& \frac{\sqrt2}{(n^2-4)^{1/4}}
\end{matrix}\right)\,.
\end{equation}
This results in the monodromy matching \eqref{hyperbolic conj classes}:
\begin{equation}
\mathfrak M_{S^1} \to g^{-1} \,\mathfrak M_{S^1}\, g = \mathfrak M(n) \,.
\end{equation}
Notice that this redefinition does not affect the embedding tensor resulting from Scherk-Schwarz-reduction.
Indeed, the $D=4$ gauged supergravity obtained upon truncation is blind to the choice of SL$(2,\mathbb Z)$ conjugacy class of the monodromy.

Interestingly, the fact that $\mathfrak{M}_{S^1}$ is in the hyperbolic conjugacy class of SL(2) also means that we can find a global parameterization of the SL(2)/SO(2) axio-dilaton coset representatives such that no compensating local SO(2) transformation on the IIB fermions is induced by the action of $\mathfrak M_{S^1}$.
The standard parameterization of $m_{\alpha\beta}$ in \eqref{standard axiodil param} can be obtained for instance from the SL(2)/SO(2) coset representative $\ell(C_0,\Phi)$ as
\begin{equation}
m_{\alpha\beta} = (\ell \ell^T)_{\alpha\beta}\,,\qquad
\ell(C_0,\Phi)\equiv
\begin{pmatrix}
e^{-\Phi/2} & -e^{\Phi/2} C_0 \\
0 & e^{\Phi/2}	
\end{pmatrix}\,,
\end{equation}
while in order to avoid SO(2) compensating transformations under $\mathfrak{M}_{S^1}$ we may for instance change parameterization to
\begin{equation}
m_{\alpha\beta} = (\ell' \ell'^T)_{\alpha\beta}\,,\qquad
\ell \to \ell' \equiv
g^{-1}\,\frac{1}{\sqrt2}\,\begin{pmatrix}	1&-1\\1&1 \end{pmatrix}\,\ell(\chi_0,\phi)\,,
\end{equation}
where an expression for the axio-dilaton in terms of $\chi_0,\phi$ can be easily constructed.
This is still a global choice of coset representative and SO(2) gauge, and now $\mathfrak{M}_{S^1}$ acts as a shift of the field $\phi$ without inducing local SO(2) transformations.

The importance of this observation lies in the fact that under a Scherk-Schwarz reduction of ExFT fermions behave as scalar densities, at least for a certain SU(8) gauge choice \cite{Hohm:2014qga}.
Hence, their dependence on internal coordinates is entirely encoded in the function $\rho$ of \eqref{UU}.
This also applies to the uplift of the gauged supergravity residual Killing spinors at the vacuum.
Because we can find an SO(2) gauge such that the $S^1$ monodromy does not require compensating gauge transformations on the fermions, we can conclude that the $\mathcal N=4$ Killing spinors of the AdS$_4$  solution described above uplift to globally well-defined Killing spinors of Type IIB.
This means that the S-fold solution above preserves 16 supercharges which are single-valued and in fact $\eta$-independent at least in an appropriate gauge.

As an aside, it is interesting to note that choosing a different $A(\eta)$ taking values in the SO(2) or $\mathbb R$ subgroups of SL(2) one arrives at an S-fold interpretation of the reduction ansatz \eqref{twistU} for the $(\mathrm{SO(6)\times SO(2)})\ltimes T^{12}$ and $(\mathrm{SO(6)\times \mathbb R})\ltimes T^{12}$ gaugings, respectively.
The $(\mathrm{SO(6)\times \mathbb R})\ltimes T^{12}$ case has a second interesting interpretation: the $\mathbb R$ valued $A(\eta)$ matrix can be interpreted as inducing $F_1={\rm d}C_0$ flux along $S^1$, while $S^5$ is supported by $F_5$.
If we $T$-dualize along $\eta$, $F_1$ goes into the Romans mass $F_0$ and $F_5$ goes into $F_6$ filling $S^5\times S^1$.
The reduction ansatz can then be reinterpreted as Type IIA on $S^5\times S^1$ with $F_6$ and $F_0$ flux, where $A(\eta)\in\mathbb R$ generates the Romans mass in terms of a linear dependence of $C_1$ on the winding coordinate $\eta = Y_{67}$ (the physical coordinate would be $Y^{68}$).
This is analogous to \cite{Hohm:2011cp} and in fact the $A(\eta)$ part of such an ansatz matches one of the non-geometric twist-matrices that generate the Romans mass provided in \cite{Ciceri:2016dmd}.
One can alternatively implement the Romans mass directly in ten dimensions in terms of a deformation of the exceptional field theory/generalised geometry \cite{Ciceri:2016dmd,Cassani:2016ncu}, and use the CSO$(6,0,2)$ Ansatz based on $(\hat\rho,\,\hat U)$ alone to implement a geometric reduction of massive IIA to  $(\mathrm{SO(6)\times \mathbb R})\ltimes T^{12}$ gauged supergravity.

\section{Discussion}

In this paper we have constructed the twist matrices that define the consistent truncation of E$_{7(7)}$
exceptional field theory down to the $D=4$ dyonic gaugings with gauge group
$({\rm SO}(p,q)\times {\rm SO}(p',q')) \ltimes N$. The twist matrix satisfies the section constraints
so that the corresponding dyonic models can be embedded either in
Type IIA ($p+q$ odd) or in Type IIB ($p+q$ even) theories. Using the dictionary between
exceptional field theory and IIB supergravity, we have worked out the explicit uplift formulas for the
$(\mathrm{SO(6)\times \SO(1,1)})\ltimes T^{12}$ gauging and given the uplift of
the four-dimensional AdS$_4$ ${\cal N}=4$ vacuum~\cite{Gallerati:2014xra}
into a supersymmetric AdS$_4\times M_5\times S^1$ S-fold solution of IIB supergravity. The internal space
$M_5$ is a deformation of the round sphere preserving an ${\rm SO}(4)\subset {\rm SO}(6)$
subset of its isometries.

Before compactification of the $\eta$ direction, the solution we construct in section~\ref{sec:solution} has the same topology as AdS$_5\times S^5$.
The parameterization we give is in the form of a warped product AdS$_4\times S^2\times S^2\times\Sigma$, where $\Sigma$ is an infinite strip parameterized by $\eta$ and $r$. At the boundary of the strip ($r=0,1$) one of the two $S^2$ smoothly shrinks to zero size, reproducing the $S^5$ topology.
Another important observation is that with a constant SL$(2,\mathbb R)$ rotation the axion can be set to vanish, while the dilaton runs along the $\eta$ and $r$ directions as
\begin{equation}
	e^{\Phi} = \frac{e^{-2\eta}}{\sqrt3}\left(\frac{3-2r^2}{1+2r^2}\right)^{1/2} \, .
\end{equation}
This strongly suggests that our solution be part of the class of Janus solutions with 16 supercharges of \cite{DHoker:2007zhm,DHoker:2007hhe}. This is indeed proven in Appendix \ref{janusmatch}.
More specifically, it corresponds to a smooth solution without NS5 or D5 sources, with the dilaton varying from $-\infty$ to $+\infty$ along the infinite stripe.
This differs from the regular Janus solution of \cite{DHoker:2007zhm,DHoker:2007hhe}, where the dilaton varies between finite boundary values.
Janus configurations and their relation with interface $\mathcal N=4$ super Yang--Mills have been largely studied in the literature \cite{Clark:2004sb,Clark:2005te,DHoker:2006qeo,Gaiotto:2008sd}.
It would be interesting to understand whether the S-fold compactified AdS$_4$ solution we find upon imposing periodicity in $\eta$ is also part of other constructions relating supersymmetric Janus solutions to three-dimensional $\mathcal N=4$ conformal field theories \cite{Gaiotto:2008ak,Assel:2011xz,Assel:2012cj,Ganor:2014pha}.
In fact, imposing periodicity in $\eta$ corresponds to compactifying the infinite strip $\Sigma$ to a finite cylinder, which seems analogous to the construction in \cite{Assel:2012cj}.  

There has also been some recent activity on S-folds in the context of $D=4$ $\mathcal N=3$ conformal field theories \cite{Aharony:2015oyb,Garcia-Etxebarria:2015wns,Aharony:2016kai}.
In those cases a generalization of the O3 orientifold projections is introduced, that acts with a $\mathbb Z_k \subset \mathrm{SL}(2,\mathbb Z)$ on the Type IIB fields and on the stack of D3 branes defining the CFT ($k=2,3,4,6$).
No dimensional reduction is performed, and the theories obtained from D3 branes on top of such background are either $\mathcal N=4$ or genuinely $\mathcal N=3$.
Only the elliptic subgroups of SL$(2,\mathbb Z)$ are used in that case, as there must be a fixed valued of the complex coupling $\tau$, so that the projection is by a symmetry of the original theory.

A distinguished property of our solution is that it arises from a consistent truncation of Type IIB supergravity to $D=4$, $(\mathrm{SO}(6)\times\mathrm{SO(1,1)})\ltimes T^{12}$ gauged maximal supergravity.
Thanks to the Scherk-Schwarz ansatz \eqref{SSansatz},  we have access to the full configuration space of the consistent truncation, which is part of the configuration space of IIB supergravity, also away from the solution with 16 supercharges. In the holographic context this gives access  also on the field theory side to a consistent truncation to a subset of operators.
On the gravity side, this can be used to generate other interesting solutions.
For instance,  other vacua of the gauged supergravity may have $\mathcal N<3$ supersymmetry\footnote{Some unstable vacua are known \cite{DallAgata:2011aa}.}
and  lift to less supersymmetric Janus solutions and their compactifications.
All types of solutions of this gauged supergravity (domain walls, black holes, etc.) now also admit a Type IIB embedding.
It would thus be very interesting to further clarify the relation of $(\mathrm{SO}(6)\times\mathrm{SO(1,1)})\ltimes T^{12}$ gauged maximal supergravity to Janus solutions with (SL(2) duality twists), and thus their relation to interface $\mathcal N=4$ super Yang--Mills and $\mathcal N=4,\ D=3$ conformal field theories.

\subsection*{Acknowledgements}
We would like to thank Franz Ciceri, Gianguido Dall'Agata, Adolfo Guarino, Emanuel Malek for helpful discussions.
We thank the Galileo Galilei Institute for Theoretical Physics for the hospitality and the
INFN for partial support during the completion of this work.
G.I. is partially supported by FCT/Portugal through a CAMGSD post-doc fellowship.

\section*{Appendix}

\begin{appendix}

\section{The ${\cal N}=4$ vacuum in moduli space}
\label{app:vacuumM}

The non-vanishing entries of $M_{MN}$ at the vacuum \eqref{vacuum coset} are
\allowdisplaybreaks
\begin{align}
M^{\underline{ij}\,\underline{kl}} &= M_{\underline{ij}\,\underline{kl}} =
\begin{cases}
3\, \delta^{\underline i [\underline k}\delta^{\underline l] \underline j}
& \underline{i},\underline{j} = 1,2,3 \\
3\, \delta^{\underline i [\underline k}\delta^{\underline l] \underline j}
& \underline{i},\underline{j} = 4,5,8 \\
\delta^{\underline i [\underline k}\delta^{\underline l] \underline j}
&\text{otherwise}
\end{cases}
\label{a1}\\[1ex]
M^{\underline{ij}}{}_{\underline{ka}} &=
\begin{cases}
-{3^{-1/4}}\, \epsilon^{\underline{ijk}}
& \underline{i},\underline{j},\underline{k} = 1,2,3\ (\epsilon^{123}=+1) \\
(-)^{\underline a+1}\, {3^{1/4}}\, \epsilon^{\underline{ijk}}
& \underline{i},\underline{j},\underline{k} = 4,5,8\ (\epsilon^{458}=+1)
\end{cases}
\label{a2}
 \\[1ex]
M^{\underline{ka}}{}_{\underline{ij}} &=
\begin{cases}
-{3^{1/4}}\, \epsilon^{\underline{ijk}}
& \underline{i},\underline{j},\underline{k} = 1,2,3\ (\epsilon^{123}=+1) \\
(-)^{\underline a+1}\, {3^{-1/4}}\, \epsilon^{\underline{ijk}}
& \underline{i},\underline{j},\underline{k} = 4,5,8\ (\epsilon^{458}=+1)
\end{cases} \\[1ex]
M_{\underline{ia}\,\underline{jb}} &=
\begin{cases}
\tfrac{\sqrt3}{2}\, \delta_{\underline{ij}} \delta_{\underline{ab}}	
& \underline{i},\underline{j} = 1,2,3 \\
\tfrac1{2\sqrt3}\, \delta_{\underline{ij}} (5\delta_{\underline{ab}}-4\sigma_{1\,\underline{ab}})
& \underline{i},\underline{j} = 4,5,8 \\
\end{cases} \\[1ex]
M^{\underline{ia}\,\underline{jb}} &=
\begin{cases}
\tfrac1{2\sqrt3}\, \delta^{\underline{ij}}  (5\delta^{\underline{ab}}+4\sigma_1^{\underline{ab}})
& \underline{i},\underline{j} = 1,2,3 \\
\tfrac{\sqrt3}{2}\, \delta^{\underline{ij}}  \delta^{\underline{ab}}	
& \underline{i},\underline{j} = 4,5,8 \\
\end{cases}\\[1ex]
M_{\underline{67}\,\underline{67}} &= M^{\underline{67}\,\underline{67}} = \delta_{6 [6}\delta_{7] 7} =1/2 \ .
\label{a6}
\end{align}

\section{E$_{7(7)}$ generators}
\label{app:generators}
\begin{align}
[T_A{}^B]_C{}^D &= 4\delta_A^C\delta_D^B -\frac12 \delta_A^B\delta_C^D\,, \\[1ex]
[T_A{}^B]_M{}^N &=
\begin{pmatrix}
    2\delta_{[C}^{[E} [T_A{}^B]^{\vphantom{[]}}_{D]}{}_{\vphantom{[]}}^{F]} & \\
 & -2\delta_{[E}^{[C} [T_A{}^B]^{\vphantom{[]}}_{F]}{}_{\vphantom{[]}}^{D]}
\end{pmatrix} \,, \\[1ex]
[T_{ABCD}]_M{}^N &=
\begin{pmatrix}
 & \varepsilon_{ABCDEFGH}	\\
 4!\,\delta_{ABCD}^{EFGH} &
\end{pmatrix} \,.
\label{eq:generators}
\end{align}
\section{Relation to the $\mathcal{N}=4$ Janus solution }\label{janusmatch}
In this appendix we show that the solution discussed in Section \ref{sec:solution}, upon suitable redefinitions and an S-duality rotation, coincides with the $\mathcal{N}=4$ supersymmetric Janus solution of \cite{Assel:2011xz}.\par
Let us define the $S^2\times S^2$ sphere harmonics as
\bea
{\cal Y}_1^p &\equiv& \frac1{r}\,{\cal Y}^p\;,\qquad
{\cal Y}_2^p ~\equiv~ \frac1{\sqrt{1-r^2}}\,{\cal Z}^p
\;,
\eea
such that ${\cal Y}_1^p{\cal Y}_1^p=1={\cal Y}_2^p{\cal Y}_2^p$\,.
Then
\bea
d{\cal Y}^p &=& r\,d{\cal Y}_1^p + {\cal Y}_1^p\,dr\;,
\nonumber\\
d{\cal Z}^p &=& \sqrt{1-r^2}\,d{\cal Y}_2^p -\frac{r}{\sqrt{1-r^2}} {\cal Y}_2^p\,dr
\eea
Let us also set
\bea
r&=&{\rm sin}\,x\;.
\eea
with $0\le x \le \pi/2$. We shall define on the surface $\Sigma$ parametrized by $\eta,\,x$ the complex coordinate $z=\eta-i\,x$, with $\mathrm{Im}z=x\in[0,\frac\pi2]$.
Upon these redefinitions, the ten-dimensional IIB metric (\ref{IIBmetric}) has the form
\bea
ds^2 &=&
\Delta^3\, {\rm sin}^2x\,(1+2\,{\rm cos}^2x)\,
d{\cal Y}_1^p  d{\cal Y}_1^p
+
\Delta^3\,(1+2\,{\rm sin}^2x)\,{\rm cos}^2x\,d{\cal Y}_2^pd{\cal Y}_2^p
\nonumber\\
&&{}
+\Delta^{-1} \left(dxdx+d\eta\, d\eta \right)
+\frac12\,\Delta^{-1}\, ds^2_{{\rm AdS_4}}
\;,
\eea
with the warp factor given by
\bea
\Delta &=& \left((1+2\,{\rm sin}^2x)(1+2\,{\rm cos}^2x)\right)^{-1/4}
\;,
\eea
and the AdS$_4$ radius fixed to $r_{\rm AdS}=1$\,.

Comparing to the notation of \cite{Assel:2011xz}, in which the metric is written as:
\bea
ds^2&=& f_4^2 ds^2_{{\rm AdS}_4}
+f_1^2\,ds^2_{S_1^2}
+f_2^2\,ds^2_{S_2^2}
+4\rho^2\,dzd\bar{z}
\;,
\eea
we can make the following identifications
\bea
 f_4^8 &=& {\frac1{16}} \, \Delta^{-4}= {\frac1{16}}\,(1+2\,{\rm sin}^2x)(1+2\,{\rm cos}^2x)
 \;,
 \nonumber\\
f_1^2 &=&  \Delta^3\, {\rm sin}^2x\,(1+2\,{\rm cos}^2x)
 \;,
 \nonumber\\
f_2^2 &=&  \Delta^3\,(1+2\,{\rm sin}^2x)\,{\rm cos}^2x
\;,\nonumber\\
4\rho^2 &=& \Delta^{-1}
\;.
\eea

As explained in Section \ref{sfoldsuper}, in order to match our solution with that of \cite{Assel:2011xz}, the following S-duality transformation has to be performed on the ${\rm SL}(2)$-covariant fields:
\begin{align}
m_{\alpha\beta}&\rightarrow m'_{\sigma\gamma}=S_\alpha{}^\sigma\,S_\beta{}^\gamma\,m_{\sigma\gamma}=\left(\begin{matrix}e^{-\Phi'} & 0 \cr 0 &e^{\Phi'} \end{matrix}\right)\,,\nonumber\\
H_3^\alpha &\rightarrow H_3^{\prime \alpha}= {S^{-1}}{}_\beta{}^\alpha\,H_3^\beta\,,
\end{align} 
where 
\bea
S_\alpha{}^\beta
&\equiv&\frac1{\sqrt{2}}\,
\begin{pmatrix}
1&-1\\
1&1
\end{pmatrix}
\;.
\eea
We then find:
\bea
m_{\alpha\beta} &=&
\begin{pmatrix}
 \sqrt{3}\,e^{2\eta}\,\frac{(1+2\,{\rm sin}^2x)^{1/2}}{(1+2\,{\rm cos}^2x)^{1/2}}& 0\\
0 & \frac1{\sqrt{3}}\,e^{-2\eta}\,\frac{(1+2\,{\rm cos}^2x)^{1/2}}{(1+2\,{\rm sin}^2x)^{1/2}}
\end{pmatrix}
\;.
\eea
From which we read off
\bea
e^{-2\Phi'} &=&
3\,e^{4\eta}\,\frac{(1+2\,{\rm sin}^2x)}{(1+2\,{\rm cos}^2x)}\,.
\;.
\eea

The three-form field strengths  take the form
\bea
H_3{}^{\prime +} &=&
\frac{\sqrt{2}\,3^{-1/4}\,e^{-\eta}\,{\rm sin}^2x}{1+2\,{\rm sin}^2x} \,\varepsilon_{pqr}
\,{\cal Y}_1^p\,d{\cal Y}_1^q\wedge d{\cal Y}_1^r\wedge
\left(
\frac{3+2\,{\rm sin}^2x}{1+2\,{\rm sin}^2x} \, {\rm cos}\,x\,dx-  {\rm sin}\,x\,d\eta\right)
\nonumber\\
&=&
\omega_{S^1_1}\wedge
db^1
\nonumber\\
H_3{}^{\prime -} &=&
 \frac{\sqrt{2}\,3^{1/4}\,e^{\eta}\,{\rm cos}^2x}{1+2\,{\rm cos}^2x}
\,\varepsilon_{pqr}\,{\cal Y}_2^p\,d{\cal Y}_2^q\wedge d{\cal Y}_2^r\wedge
\left(
\frac{3+2\,{\rm cos}^2x}{1+2\,{\rm cos}^2x}\,
 {\rm sin}\,x\,dx -{\rm cos}\,x\,d\eta\right)
 \nonumber\\
 &=&
 \omega_{S^1_2}\wedge
db^2
\;.
\eea
where
\begin{align}
b^1&=\frac{2\sqrt{2}\,3^{-1/4}\,e^{-\eta}\,{\rm sin}^3x}{1+2\,{\rm sin}^2x}\,,\qquad
b^2=-\frac{2\sqrt{2}\,3^{1/4}\,e^{\eta}\,{\rm cos}^3x}{1+2\,{\rm cos}^2x}\,.
\end{align}
Finally, the selfdual IIB five-form field strength is given by
\bea
H_5 &=& \frac94\,\Delta^4\,{\rm sin}^2x\,{\rm cos}^2x
\,{\cal Y}_2^{p}\,
{\cal Y}_1^{[p} d{\cal Y}_1^q\wedge d{\cal Y}_1^{r]}\wedge  d{\cal Y}_2^{q}\wedge d{\cal Y}_2^r\wedge
\left(dx+\frac43\,{\rm sin}\,x\,{\rm cos}\,x  \,d\eta\right)
\nonumber\\
&&{}
-\frac1{16\cdot\, {4}}\,\sqrt{|g|}\,\varepsilon_{\mu\nu\rho\sigma} \, dx^\mu \wedge dx^\nu \wedge dx^\rho \wedge dx^\sigma \wedge
\left(
d\eta  -\frac43\, {\rm sin}\,x\,{\rm cos}\,x \, dx
\right)
\nonumber\\
&=& \frac32\,\Delta^4\,{\rm sin}^2x\,{\rm cos}^2x
\,\omega_{S^1_1}\wedge \omega_{S^1_2}
\wedge
\left(dx+\frac43\,{\rm sin}\,x\,{\rm cos}\,x  \,d\eta\right)
\nonumber\\
&&{}
-\frac{3}{2\cdot\, {4}}\,\omega_{0123} \wedge
\left(
d\eta  -\frac43\, {\rm sin}\,x\,{\rm cos}\,x \, dx
\right)
\;.\label{5formj}
\eea
where we used the property:
\bea
{\cal Y}_1^{[p} d{\cal Y}_1^q\wedge d{\cal Y}_1^{r]} &=& \frac13\,\varepsilon^{pqr}\,\omega_{S^1}
\;.
\eea
To compare the solution to the Janus one, is is useful to write $H_5$ in the form:
\bea
H_5 &=& f_1^2f_2^2\,
\,\omega_{S^1_1}\wedge \omega_{S^1_2}
\wedge
\left(\ast_2{\cal F}\right)
-f_4^4\,\omega_{0123} \wedge
{\cal F}
\;,
\eea
where $\ast_2$ is the Hodge duality operation on the disk spanned by $\eta$ and $x$, and
\bea
f_4^4\,{\cal F} &=& dj_1\,\,,\,\,\,\,j_1\equiv \frac18\,(3\eta+{\rm cos}(2x))\,.
\;,
\eea
so that
\bea
f_4^4\,{\cal F} &=& \frac38 \left(d\eta-\frac43\,{\rm cos}\,x\,{\rm sin}\,x\,dx\right)
\;,
\eea
and 
\bea
f_1^2f_2^2\,\ast_2\!{\cal F} &=&  \frac32\,\Delta^4\,{\rm sin}^2x\,{\rm cos}^2x \left(dx+\frac43\,{\rm cos}\,x\,{\rm sin}\,x\,d\eta\right)
\eea

\subsection{Reconstruct the solution from the harmonic functions $\mathcal{A}_1$, $\mathcal{A}_2$}
To show the matching of the above solution with that in \cite{Assel:2011xz} we need to prove that all the functions describing it can be expressed, through appropriate relations  given in the same reference, in terms of only two harmonic functions $\mathcal{A}_1,\,\mathcal{A}_2$ on the surface $\Sigma$ spanned by $\eta,\,x$, or, equivalently, by the complex coordinate $z=\eta-i\,x$.
This readily follows from the identification
\bea
{\cal A}_1 &=& \frac{3^{1/4}}{4\sqrt{2}}\,e^{z}  \;,\qquad
{\cal A}_2~=~ -\frac{3^{-1/4}}{4\sqrt{2}}\,e^{-z}\;,
\eea
in terms of which we define the harmonic functions $h_1,\,h_2$ and their duals $\tilde{h}_1,\,\tilde{h}_1$:
\bea
h_1 &=& -i({\cal A}_1-\bar{\cal A}_1)~=~
-\frac{3^{1/4}}{2\sqrt{2}}\,e^\eta\,{\rm sin}\,x
\;,\nonumber\\
h_2 &=&{\cal A}_2+\bar{\cal A}_2~=~
-\frac{3^{-1/4}}{2\sqrt{2}}\,e^{-\eta}\,{\rm cos}\,x
\;,
\eea
\bea
\tilde{h}_1&=&{\cal A}_1+\bar{\cal A}_1~=~
\frac{3^{1/4}}{2\sqrt{2}}\,e^\eta\,{\rm cos}\,x
\;,\nonumber\\
\tilde{h}_2&=&i({\cal A}_2-\bar{\cal A}_2)~=~
\frac{3^{-1/4}}{2\sqrt{2}}\,e^{-\eta}\,{\rm sin}\,x
\;.
\eea
It is then straightforward to show that the functions entering the solution satisfy the following relations characterizing the solution of \cite{Assel:2011xz}:
\bea
W&=&\partial h_1 \bar\partial h_2+\bar\partial h_1 \partial h_2
~=~-\frac18\,{\rm sin}\,x\,{\rm cos}\,x
\;,\nonumber\\
N_1 &=&
2\,h_1h_2\,|\partial h_1|^2-h_1^2\,W ~=~
\frac{\sqrt{3}}{128}\,e^{2\eta}\,
{\rm sin}\,x\,{\rm cos}\,x\,(1+2\,{\rm sin}^2 x)
\;,\nonumber\\
N_2 &=&
2\,h_1h_2\,|\partial h_2|^2-h_2^2\,W ~=~
\frac{1}{\sqrt{3}\,128}\,e^{-2\eta}\,
{\rm sin}\,x\,{\rm cos}\,x\,(1+2\,{\rm cos}^2 x)
\;,\nonumber\\
f_4^8 &=& 16\,\frac{N_1N_2}{W^2}~=~\frac1{16}\,\Delta^{-4}
\;,\nonumber\\
(4\,\rho^2)^4 &=&
256\,\frac{N_1N_2W^2}{h_1^4h_2^4}~=~\Delta^{-4}
\;,\nonumber\\
f_1^8 &=& 16\,h_1^8\,\frac{N_2\,W^2}{N_1^3}~=~
{\rm sin}^8 x \,(1+2\,{\rm cos}^2x)(1+2\,{\rm sin}^2x)^{-3}
\;,\nonumber\\
f_2^8 &=& 16\,h_2^8\,\frac{N_1\,W^2}{N_2^3}~=~
{\rm cos}^8 x \,(1+2\,{\rm sin}^2x)(1+2\,{\rm cos}^2x)^{-3}
\;,\nonumber\\
e^{-2\Phi'} &=&\frac{N_1}{N_2}=
3\,e^{4\eta}\,\frac{(1+2\,{\rm sin}^2x)}{(1+2\,{\rm cos}^2x)}\,.
\;.
\eea
We also find that the two functions $b_1,\,b_2$ entering the expression of the 3-form field strengths are related to the above functions as prescribed in \cite{Assel:2011xz}
\bea
b_1 &=&2i \frac{h_1}{N_1}\,h_1h_2\,(\partial h_1\bar\partial h_2 - \bar\partial h_1 \partial h_2)
+2\tilde{h}_2
~=~ \frac{2\sqrt{2}\cdot3^{-1/4}\, e^{-\eta}}{1+2\,{\rm sin}^2x}\,{\rm sin}^3 x\;,
\nonumber\\
b_2 &=&2i \frac{h_2}{N_2}\,h_1h_2\,(\partial h_1\bar\partial h_2 - \bar\partial h_1 \partial h_2)
-2\tilde{h}_1
~=~ -\frac{2\sqrt{2}\cdot3^{1/4} \,e^{\eta}}{1+2\,{\rm cos}^2x}\,{\rm cos}^3 x\;.
\eea
Similarly, just as in the Janus solution, the function $j_1$ entering the five-form field strength can be expressed as:
\begin{equation}
j_1=3\left(\mathcal{C}+\overline{\mathcal{C}}- \mathcal{D}\right)+\frac{i h_1 h_2}{W}\left(\partial h_1 \overline{\partial}h_2-\partial h_2 \overline{\partial}h_1\right)\,,
\end{equation}
where $\mathcal{C}$ satisfies the relation $\partial \mathcal{C}=\mathcal{A}_1 \partial \mathcal{A}_2-\mathcal{A}_2 \partial \mathcal{A}_1$ and is given by $\mathcal{C}=\frac{z}{16}$, while $\mathcal{D}$ reads 
\begin{align}
\mathcal{D}=\overline{\mathcal{A}}_1\mathcal{A}_2+\overline{\mathcal{A}}_2\mathcal{A}_1= -\frac{1}{16}\,\cos(2x)\,.
\end{align}
\end{appendix}


\providecommand{\href}[2]{#2}\begingroup\raggedright\endgroup

\end{document}